\documentclass[10pt, letterpaper]{article}
\pdfoutput=1

\usepackage{amsmath,amsfonts,amssymb,amsthm}
\usepackage{mathtools}
\usepackage[usenames,dvipsnames,svgnames,table]{xcolor}
\usepackage{thm-restate}
\usepackage{braket}
\usepackage{hyperref}
\usepackage{wrapfig}
\usepackage{enumitem}
\usepackage{tocloft}
\usepackage{microtype}
\usepackage{stmaryrd}
\usepackage[utf8]{inputenc}
\usepackage{physics}
\usepackage{graphicx}
\usepackage{setspace}
\usepackage{changepage}
\usepackage{hyphenat}
\usepackage{caption}
\usepackage{float}
\usepackage[linesnumbered,ruled,vlined]{algorithm2e}
\usepackage{setspace}
\usepackage{etoolbox}
\usepackage{tcolorbox}
\usepackage{tikz,tikz-qtree}
\usepackage{stackengine}
\stackMath

\usetikzlibrary{shapes.misc,shapes.geometric,arrows,positioning,calc,backgrounds}

\newcommand{\doverline}[1]{\stackon[1.2pt]{\widehat{#1}}{\widehat{\phantom{#1}}}}
\renewcommand\overline\widehat

\tikzstyle{bubble} = [rectangle, rounded corners, minimum width=2cm, minimum height=1cm, text centered, draw=black, fill=black!5!white, text width=2cm]
\tikzstyle{arrow} = [thick, ->, >=stealth]
\tikzstyle{gateset} = [rectangle, minimum width=2cm, text centered, minimum height=1cm, fill=white, text width=2cm]

\newcommand{\Mod}[1]{\ \mathrm{mod}\ #1}

\SetAlFnt{\small}
\SetAlCapFnt{\normalsize}
\SetAlCapNameFnt{\normalsize}

\makeatletter
\patchcmd{\@algocf@start}
  {-1.5em}
  {0pt}
  {}{}
\makeatother

\DontPrintSemicolon

\SetKwComment{Comment}{\color{green!50!black}// }{}

\newcommand{\assign}{\leftarrow}
\newcommand{\FuncCall}[2]{\texttt{\bfseries #1(#2)}}
\SetKwProg{Function}{function}{}{}

\hypersetup{
    pdftitle={Efficient Universal Quantum Compilation: An Inverse-Free Solovay-Kitaev Algorithm},
    pdfauthor={Adam Bouland, Tudor Giurgica-Tiron},
    colorlinks=true,
    linkcolor=blue!80!black,
    filecolor=blue!80!black,      
    urlcolor=blue!80!black,
    citecolor=blue!80!black
}

\newtheoremstyle{thrmstyle}
  {0pt} 
  {20pt} 
  {\em} 
  {} 
  {\bfseries} 
  {.} 
  {.5em} 
  {} 
\theoremstyle{thrmstyle}
\newtheorem{theorem}{Theorem}[section]
\newtheorem{lemma}{Lemma}[section]

\renewenvironment{proof}{\emph{\bfseries Proof.}}{\qed\\\vspace{10pt}}

\newtheorem{innercustomthm}{Theorem}
\newenvironment{customthm}[1]
  {\renewcommand\theinnercustomthm{#1}\innercustomthm}
  {\endinnercustomthm}

\usepackage[left=1in,right=1in,top=1in,bottom=1in]{geometry}
\tcbuselibrary{most}
\tcbset{colback=black!3!white,colframe=black, enhanced, boxrule=0.2mm, arc=0.75mm}

\begin{document}

\author{
Adam Bouland\\[.5ex]
{\normalsize Stanford University}\\
{\normalsize \texttt{abouland@stanford.edu}}
\and
\qquad
\and
Tudor Giurgic\u{a}-Tiron\\[.5ex]
{\normalsize Stanford University}\\
{\normalsize \texttt{tgt@stanford.edu}}
}

\date{}

\title{\vspace{-20pt}\bfseries{\Large Efficient Universal Quantum Compilation: \\ An Inverse-free Solovay-Kitaev Algorithm}}

\maketitle

\begin{abstract}

\noindent The Solovay-Kitaev algorithm is a fundamental result in quantum computation. It gives an algorithm for efficiently compiling arbitrary unitaries using universal gate sets: any unitary can be approximated by short gates sequences, whose length scales merely poly-logarithmically with accuracy. As a consequence, the choice of gate set is typically unimportant in quantum computing. However, the Solovay-Kitaev algorithm requires the gate set to be inverse-closed. It has been a longstanding open question if efficient algorithmic compilation is possible without this condition. In this work, we provide the first inverse-free Solovay-Kitaev algorithm, which makes no assumption on the structure within a gate set beyond universality, answering this problem in the affirmative, and providing an efficient compilation algorithm in the absence of inverses for both $\text{SU}(d)$ and $\text{SL}(d, \mathbb C)$. The algorithm works by showing that approximate gate implementations of the generalized Pauli group can \emph{self-correct} their errors.

\end{abstract}

\vspace{10pt}
\section{Introduction}

The Solovay-Kitaev (S-K) algorithm is a central result in quantum compilation. 
It shows how to approximate arbitrary unitary operations using elements from a finite, universal gate set. 
In particular, it gives an explicit algorithm which, given an inverse-closed universal gate set $\mathcal G$ and a target unitary $U$, $\epsilon$-approximates $U$ using merely $\text{polylog}(\epsilon^{-1})$ gates from $\mathcal G$ \cite{kitaev1997quantum,dawson2005}. The Solovay-Kitaev algorithm guarantees that the runtime of quantum algorithms is independent of the choice of gate set up to poly-logarithmic factors. This is because any $m$-gate algorithm in one gate set can always be compiled to a circuit of size $O(m\,\text{polylog}\, m)$ in another gate set\footnote{This is because $O(m^{-1})$ compilation error per gate suffices for any $\mathsf{BQP}$ algorithm.}.
Conversely, if compilation instead required $O(\epsilon^{-1})$ overhead, then implementation in another gate set could quadratically worsen runtimes, which would severely restrict the practicality of polynomial speedups such as Grover search.
Given its central importance to both theoretical and practical compilation, there have been many subsequent improvements to the S-K algorithm, for example developing highly efficient compilation algorithms for particular gate sets \cite{ ross2014optimal, forest2015exact, bocharov2015efficient,kliuchnikov2015practical,parzanchevski2018super}.\\

One limitation of the standard Solovay-Kitaev theorem is that it requires the gate set to be {\em inverse-closed}: for every element in the gate set, its exact inverse is also available. Mathematically, this is extremely convenient: the S-K algorithm is based on implementing group commutators made of gate sequences and their inverses. With an inverse-closed gate set, one can natively implement the inverse of a sequence of gates by inverting each gate element individually and reversing the order.
However, it has long been an open problem if the restriction to inverse-closed gate sets is necessary, and whether an inverse-free Solovay-Kitaev algorithm exists with similar poly-logarithmic asymptotics. 
This question is presented as a longstanding problem in Dawson and Nielsen \cite{dawson2005} and Kuperberg \cite{kuperberg2009hard}.
To the best of our knowledge, even special cases of inverse-free S-K --- for example when the gate set consists of two rotations of the Bloch sphere around non-collinear axes by irrational multiples of $\pi$ --- were open problems.\\

Prior work has made partial progress on removing the inverse-closure condition, but at the cost of requiring other forms of exact mathematical structure in the gate set.
First, Sardharwalla, Cubitt, Harrow, and Linden \cite{sardharwalla2016universal} showed that if the gate set contains an exact implementation of the generalized Pauli (also known as the Weyl) group, then an efficient S-K algorithm is possible. 
Shortly thereafter, Bouland and Ozols \cite{bouland2017trading} generalized this to the case when the gate set contains an exact projective irreducible representation (irrep) of \emph{any} finite group. 
However, neither work removed the need for some form of exact structure in the gate set --- in this case inverse-closure is traded for an irrep. 
Interestingly, Oszmaniec, Sawicki, and Horodecki recently proved an \emph{information-theoretic} version of S-K without inverses \cite{oszmaniec2020epsilon} --- in other words, there \emph{exist} short sequences of gates to approximate any unitary  --- generalizing \cite{harrow2002efficient} and using ideas from the mathematics literature on random walks in compact groups \cite{bourgain2012spectral, varju2013random}.
However this left open the problem of \emph{algorithmically finding} those short sequences, as their proof is nonconstructive.\\

In this work, we give the first inverse-free Solovay-Kitaev algorithm, answering the open question in the affirmative. Our algorithm efficiently finds short sequences which approximate arbitrary unitaries in $\text{SU}(d)$ with any universal gate set --- while making no assumptions about the mathematical structure within the gate set:
\begin{adjustwidth}{1.25cm}{1.25cm}
\begin{customthm}{}[Inverse-free Solovay-Kitaev Algorithm]\label{thrm:ifsk}
    For any fixed dimension $d\geq 2$, and any gate set which densely generates $\text{SU}(d)$, there is an efficient algorithm which approximates any unitary $U \in \text{SU}(d)$ within $\epsilon$ error in operator norm, using merely $O\left(\text{polylog}\,(\epsilon^{-1})\right)$ elements from the gate set.\\
\end{customthm}
\end{adjustwidth}

In our algorithm (as in \cite{sardharwalla2016universal,bouland2017trading}), the exponent of the logarithm depends on $d$, but only logarithmically so. For example in the case of a qubit the exponent is roughly $8.62$, compared to an exponent of roughly $3.97$ for the standard S-K algorithm with inverses.
We also show this algorithm easily extends to the group $\text{SL}(d, \mathbb C)$ as well, building on prior work of \cite{aharonov2007polynomial,kuperberg2009hard}.\\

\vspace{-1em}
\subsection{Proof Sketch}

Our algorithm follows the standard Solovay-Kitaev recursion: at every step, the approximation error drops from $\epsilon$ to $O(\epsilon^{3/2})$, while the sequence length is multiplied by a constant factor. 
Upon recursion, the errors drop doubly exponentially while the sequence lengths grow singly exponentially, resulting in the poly-logarithmic scaling of length with error. 
In the standard S-K algorithm, this improved approximation error arises from constructing a certain {\em group commutator} of the form $VWV^\dagger W^\dagger$.
The structure of the Lie algebra of the special unitary group allows this group commutator to polynomially suppress errors.
For this construction to be meaningful, one needs $V,W,V^\dagger$ and $W^\dagger$ to all have expressions as products of gates, and crucially that $VV^\dagger=WW^\dagger=I$ exactly, in other words $V,V^\dagger$ and $W, W^\dagger$ are pairs of \emph{exact inverses}.
If the gate set is inverse-closed, then gate implementations to $V,W$ automatically imply gate implementations of the exact inverses $V^\dagger,W^\dagger$ --- one can simply invert a gate sequence for $V$ gate by gate to obtain a gate sequence for $V^\dagger$.
Without inverse closure, it is unclear how to proceed.\\
 
 Recently, \cite{sardharwalla2016universal, bouland2017trading} noticed that the inverses in the group commutator do not need to be exact; it suffices to have the capacity to produce operator inverses to $O(\epsilon^2)$ error. The two works also proved that an exact group irrep in the gate set can be used to construct such an {\em inverse factory} by twirling over the group elements. The remaining assumption is that such group elements need to be exact themselves.
This of course requires the gate set to contain certain exact inverses --- namely those in the finite group irrep --- but not others.
It would be natural to attempt to naively obtain an inverse-free S-K by simply plugging in $\epsilon$-approximate implementations of the group elements into the group twirl.
Unfortunately this basic approximate group twirl does not work --- i.e. performing a group twirl by an approximate irrep does not allow one to create a precise enough inverse factory, as the errors in the group elements are not themselves cancelled in the twirl.
Therefore new ideas are required to make this sort of algorithm work.\\

In our work, we remove the need for any exact group structure in two steps.
First, we show that an approximate irrep of the generalized Pauli group can form \emph{self-correcting sequences} --- specifically, $O(\epsilon)$-approximate implementations of Pauli operators can form sequences which come within $O(\epsilon^2)$ of the identity. Second, we then show that the existence of self-correcting sequences allows us to create ``inverse factories'' to $O(\epsilon^2)$ precision for arbitrary unitaries, which can be used in an inverse-free Solovay-Kitaev algorithm.\\

To see this in more concrete detail, consider the case of a qubit ($d=2$) and the standard Pauli group. 
Suppose one has $\epsilon$-approximate gate implementations of $X,Z$, and call these $X',Z'$.
Then one can show that
\begin{equation*}
    Z'\,X'\,X'\,Z'\,X'\,Z'\,Z'\,X' = I + O(\epsilon^2)\,,
\end{equation*}
regardless of the specific direction of each error in $X'$ and $Z'$ (which are in general independent of one another). We call such a sequence a \emph{self-correcting sequence}, or alternatively a \emph{self-focusing sequence} by analogy with the terminology in \cite{sardharwalla2016universal}.
At first, it may seem counterintuitive that such objects exist --- as one must somehow cancel out independent errors in both the $X'$ and $Z'$ operators simultaneously to leading order --- and we will describe their mechanism of action shortly in Sections \ref{subsec:sequenceintuition} and \ref{subsec:sequenceintuition2} below.\\

Next, we generalize this construction to find self-correcting sequences for arbitrary dimensions $d\geq 2$. For example, we show that
\begin{equation}
\label{eq:selfcorrectingexample}
    \left[Z'\,X'^d\right]^{d-1}Z'\,\left[X'\,Z'^d\right]^{d-1}X' = I + O(\epsilon^2)\,
\end{equation}
is a self-correcting sequence for any $d\geq 2$, where $X'$ and $Z'$ are $\epsilon$-approximations to the $d$-dimensional generalized Pauli $X$ (i.e. the shift operator) and $Z$ (i.e. the clock operator by the $d$th root of unity).
We note that these self-correcting sequences are not unique --- though we note our self-correcting sequence for a qubit is minimal, which can be shown by a brute force search. For our purposes it suffices to find a single example of such a sequence for each $d$.\\

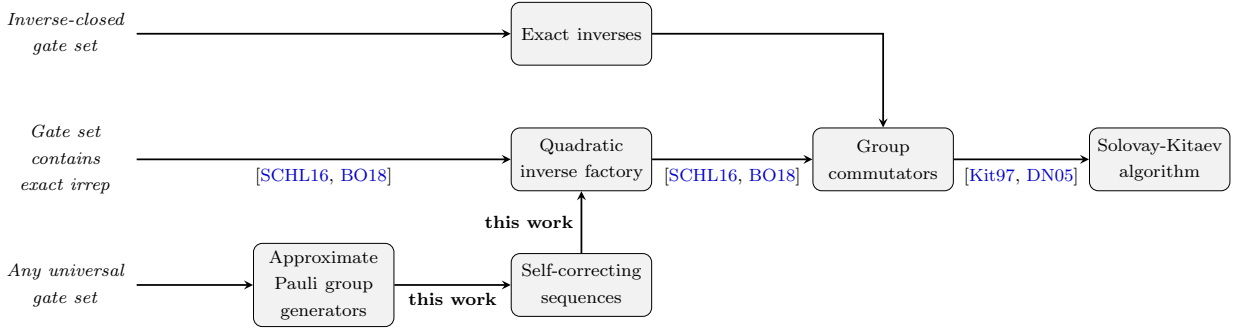
\begin{figure}
\makebox[\textwidth][c]{
\resizebox{\textwidth}{!}{
\begin{tikzpicture}[node distance=3.5cm]
    
    \node (SK) [bubble] {\footnotesize Solovay-Kitaev algorithm};
    \node (GC) [bubble, left of=SK, xshift=-0.9cm] {\footnotesize Group commutators};
    \node (IF) [bubble, left of=GC, xshift=-1.3cm] {\footnotesize Quadratic inverse factory};
    \node (EI) [bubble, above of=IF, yshift=-1.5cm] {\footnotesize Exact inverses};
    \node (SCI) [bubble, below of=IF, yshift=1.5cm] {\footnotesize Self-correcting sequences};
    \node (AG) [bubble, left of=SCI, xshift=-0.6cm] {\footnotesize Approximate Pauli group generators};
    
    \node (AnyGS) [gateset, left of=AG, xshift=-0.6cm] {\footnotesize \nohyphens{\emph{Any universal gate set}}};
    
    \node (IrrepGS) [gateset, above of=AnyGS, yshift=-1.5cm] {\footnotesize \nohyphens{\emph{Gate set contains exact irrep}}};
    
    \node (ICGS) [gateset, above of=IrrepGS, yshift=-1.5cm] {\footnotesize \nohyphens{\emph{Inverse-closed gate set}}};
    
    \draw [arrow] (GC) -- node[anchor=north] {\footnotesize \cite{kitaev1997quantum, dawson2005}}(SK);
    \draw [arrow] (IF) -- node[anchor=north, align=center] {\footnotesize \cite{sardharwalla2016universal, bouland2017trading}} (GC);
    \draw [arrow] (EI) -| (GC);
    \draw [arrow] (SCI) -- node[anchor=east] {\footnotesize \bfseries{this work}} (IF);
    \draw [arrow] (AG) -- node[anchor=north] {\footnotesize \bfseries{this work}} (SCI);
    \draw [arrow] (AnyGS) -- (AG);
    \draw [arrow] (IrrepGS) -- node[anchor=north] {\footnotesize \cite{sardharwalla2016universal,bouland2017trading}} (IF);
    \draw [arrow] (ICGS) -- (EI);
    
\end{tikzpicture}}}
\caption{An outline of the main ingredients that go into our inverse-free Solovay-Kitaev algorithm, and where they fit in relation to prior works. From top to bottom, the starting points involve gate sets with progressively fewer assumptions, ending with the structure-agnostic universal gate sets studied in this work.}
\label{fig:flowchart}
\end{figure}

We then show that self-correcting sequences are algorithmically useful, because they can be used to construct more precise inverses of arbitrary operators.
For example, one can easily see that the sequence
\begin{equation*}
      Z'\,X'\,Z'\,Z'\,X'\,Z'\,X' = X'^\dagger + O(\epsilon^2)\,
 \end{equation*}
 is an $O(\epsilon^2)$ gate approximation to $X'^\dagger$, by peeling off an instance of $X'$ from the left of the self-correcting sequence above in the case of a qubit. While this sort of argument immediately produces precise inverses of approximate Pauli elements, we show how to extend this idea to construct an ``inverse factory'' for arbitrary unitaries based on a Pauli self-correcting sequence.
 Specifically, given a unitary $V$ as a gate implementation, and an $\epsilon$-approximate gate implementation of $V^\dagger$, call it $\overline{V^\dagger}$, and additionally $\epsilon$-approximate gate implementations $X',Z'$ of $X$ and $Z$, we can construct\footnote{The key idea is to construct a different $\epsilon$-approximation to $X$ such as $V \overline{V^\dagger} X'$ and use this as a stand-in for $X'$ in the original Pauli self-correcting sequence --- this results in a sequence close to the identity from which one can pull off a factor of $V$ from the periphery to result in a precise inverse. For more detail and explicit constructions, see lemma \ref{lemma:inversefactoryqubit} for the qubit case and lemma \ref{lemma:inversesd} for the general $d$-dimensional case.} an $O(\epsilon^2)$-approximation to $V^\dagger$.
 As in \cite{sardharwalla2016universal,bouland2017trading}, we show these approximate inverses are precise enough to be substituted in the place of exact inverses in the standard Solovay-Kitaev routine, such that the same poly-logarithmic asymptotics are maintained. The way these components are put together is illustrated in Figure \ref{fig:flowchart}.\\

Finally, we also show how our construction, developed for the special unitary group $\text{SU}(d)$\footnote{One might have noticed that the Pauli generators $X$ and $Z$ are not in $\text{SU}(d)$ in even dimensions, since they have determinant $-1$. This is not an issue since we can simply multiply by the appropriate global phase to restore unit determinant, and we are left with a projective irrep instead of a standard irrep, for which all methods in this paper apply as noted in \cite{bouland2017trading}.}, can naturally extend to the larger special linear group $\text{SL}(d, \mathbb C)$. An inverse-closed algorithm for $\text{SL}(d, \mathbb C)$ was introduced by Aharonov, Arad, Ebad, and Landau \cite{aharonov2007polynomial}, and a more general version for perfect, connected Lie groups was developed by Kuperberg \cite{kuperberg2009hard}. Just like in the unitary case, the ability to engineer $\epsilon^2$-precise inverses from imperfect $\epsilon$-precise components can be used to extend these algorithms into an inverse-free Solovay-Kitaev algorithm for $\text{SL}(d, \mathbb C)$.

\subsection{The intuition behind self-correcting sequences}
\label{subsec:sequenceintuition}
As our proof hinges on the construction of self-correcting sequences for all $d$, we now provide a pedagogical explanation of their mechanism. At a technical level, the proof that these sequences are self-correcting requires only elementary algebraic manipulations. In particular, we use the fact that the generalized Pauli group also forms a linear basis for the space of $d\times d$ matrices, so one can express the errors in the implementations of Pauli operators in this basis. We find that the error in a generalized Pauli operator is quadratically suppressed in the direction of the operator itself, due to the determinant-1 property of $\text{SU}(d)$. By separately keeping track of the errors in these different directions of the Lie algebra, we show that each degree of freedom in the errors separately cancels at linear order.
For a full calculation of an example self-correcting sequence, see Section \ref{sec:SU(d)}.\\

The deeper reason that the construction works, however, comes from a combination of representation theory and combinatorics. At a high level, one can show that these self-correcting sequences essentially perform a nested group twirl over two subgroups of the generalized Pauli group --- first the subgroup generated by $X$ in the form of $\{I,X,X^2,\ldots X^{d-1}\}$, and then the similar subgroup generated by $Z$.
As noted in \cite{bouland2017trading}, twirling over a group projects errors onto the subspace preserved by the adjoint action of the group (which was trivial in their case as they worked with an exact irrep, but is nontrival for groups like $\{I,X,X^2,\ldots X^{d-1}\}$ which are not irreps).
In our case, we show that these nested group twirls cancel out \emph{complementary} directions of the errors --- leaving nothing behind at linear order, regardless of the exact direction of the $O(\epsilon)$ initial errors.
The combinatorics of the self-correcting sequences perform this nested group twirl simultaneously on the error in the $X$ operator and the error in the $Z$ operator in a nontrivial fashion.\\

\vspace{-1em}
\subsection{In more detail: cancellation by complementary group twirls}\label{subsec:selfcorrectdiscussion}
\label{subsec:sequenceintuition2}
To see this effect even more concretely, we now explain how this cancellation occurs in the self-correcting sequence \eqref{eq:selfcorrectingexample}, which works for any $\epsilon$-approximate implementations of the generalized $X$ and $Z$ operators in $d$ dimensions. Here $X$ is defined as the shift operator $X\ket{k} = \ket{k+1 \mod d}$, and $Z$ is the clock operator $Z\ket{k} = \omega^k \ket{k}$. We use the notation $\omega\equiv e^{\frac{2\pi i}{d}}$ for the $d$th root of unity. Building a self-correcting sequence out of approximate $X$ and $Z$ operators is enough for the purpose of constructing suitable inverse factories for S-K; however, this construction can generalize to any Pauli group elements, and we refer the interested reader to Section \ref{sec:SU(d)} for a particular recipe for self-correcting sequences involving arbitrary generalized Pauli operators.\\

Suppose that $X'$ is an $\epsilon$-approximate gate sequence for generalized $X$ and likewise for $Z'$.
So we can write $X'=X + \delta$ for some error $\delta$ of magnitude $\epsilon$.
Our goal is to ``cure'' these errors, i.e. take a product of the operators to reduce the errors to $O(\epsilon^2)$. First, note that if we simply take powers of $X'$, this already cures part of the error.
Namely, note that
\begin{equation}
    X'^d = (X + \delta)^d = I + \delta\,X^{d-1} + X\,\delta \,X^{d-2} + \dots + X^{d-1}\delta + O(\epsilon^2)\,.
\end{equation}
To first order this is a twirl over the group $\{I,X,X^2\ldots X^{d-1}\}$ (up to a left multiplication by $X$). Therefore, to first order this projects our error $\delta$ onto the directions preserved by the adjoint action of this subgroup\footnote{The suppression of errors in directions not preserved by the group is a consequence of Schur orthogonality, following the work of \cite{bouland2017trading}. See Section \ref{subsec:reviewirrep} for details.} --- namely in the space of operators spanned by $I,X,X^2\ldots, X^{d-1}$. 
Let us call this remaining first-order error $r$, i.e.
\[r\equiv\delta\,X^{d-1} + X\,\delta \,X^{d-2} + \dots + X^{d-1}\delta\]
We have already reduced the dimensionality of the error term from $d^2$ (the dimension of an arbitrary matrix $\delta$) to $d$ (the dimension of $r$ which is restricted to the $X$-generated subspace).
Furthermore, one can show that the component of $r$ along the identity is already suppressed, due to the Lie algebra of the special unitary group being traceless --- so in fact $r$ lies in the $(d-1)$-dimensional space spanned by $\{X,X^2\ldots X^{d-1}\}$.\\

To see how to eliminate these remaining error directions, let us first make the simplifying assumption of having access to a \emph{perfect} $Z$ operator. Consider the sequence
\begin{equation}
    \left[ZX'^d\right]^{d} = \underbrace{\left(Z X'^d\right) \left(Z X'^d\right) \dots\, \left(Z X'^d\right)}_{d \text{ times}}\,.
\end{equation}
To first order, one can easily see that the overall error is a twirl of $r$ by the $Z$-subgroup $\{I,Z,Z^2,\ldots, Z^{d-1}\}$: 
\begin{equation}
    \left[ZX'^d\right]^{d} = \left[Z\,(I + r)\right]^d = I + ZrZ^{d-1} + Z^2 r Z^{d-2} + \ldots + Z^{d} r + O(\epsilon^2)\,.
\end{equation}
This group twirl projects onto the subspace preserved by the adjoint action of the $Z$ subgroup --- which is the space spanned by $\{I,Z,Z^2,\ldots, Z^{d-1}\}$. Now note that $r$ is entirely perpendicular to that subspace --- as it lies only in the direction of the $X$ subgroup (minus the identity) --- so the error $r$ is killed the $Z$-subgroup twirl. So this means that $r$ is fully suppressed at first order, and we have that
\begin{equation}
\label{eq:xcuring}
    \left[Z\,X'^d\right]^{d} = I + O(\epsilon^2)\,,
\end{equation}
as desired. 
Symmetrically, one can show that if one had an imperfect $Z'$ and a perfect $X$ then the following analogous sequence also works\footnote{Notice that the order of the approximate and exact elements are different in the two curing sequences \eqref{eq:xcuring} and \eqref{eq:zcuring}. This is simply because we have chosen two different orderings in which to do the final group twirls. Specifically, \eqref{eq:xcuring} is obtained by twirling $X'^d$ with the ordering $\{Z,Z^2,\dots,Z^{d-1}, I\}$, while on the other hand \eqref{eq:zcuring} is obtained by twirling $Z'^d$ with the ordering $\{I,X, X^2,\dots,X^{d-1}\}$. This is a design choice made for mathematical convenience, as will become clear shortly.}:
\begin{equation}
\label{eq:zcuring}
    \left[Z'^d\,X\right]^{d} = I + O(\epsilon^2)\,.
\end{equation}

In short, if we have a perfect $X$ and an imperfect $Z$, or vice versa, we can suppress our errors at linear order. However, we need a construction that cures errors in $X$ and $Z$ simultaneously by only applying approximate $X$ and $Z$ operators. This is where the combinatorics of the self-correcting sequences come into play. The key insight is that one can find a \emph{single parent sequence} that in some sense contains \emph{both} the above sequences \eqref{eq:xcuring},\eqref{eq:zcuring} as subsequences, and therefore corrects \emph{both} the errors in $X$ and $Z$ to leading order.
More formally, we create a sequence $S$ of $X'$ and $Z'$'s which satisfies the following properties:
\begin{itemize}
    \item If $X'=X$ then $S$ simplifies to \eqref{eq:zcuring}, so $S$ cures errors in $Z$\,.
    \item If $Z'=Z$ then $S$ simplifies to \eqref{eq:xcuring}, so $S$ cures errors in $X$\,.
\end{itemize}
If we can create such a sequence, then it must be a self-correcting sequence. This is because for any sequence of $X',Z'$ operators, which to zeroth order multiples to the identity, then to first order its error is the sum of the errors induced by (a) having the incorrect $X'$ and the correct $Z$ and (b) vice versa --- simply because all cross-terms appear at sub-leading order.\\

One can show that the self-correcting sequence we provide in \eqref{eq:selfcorrectingexample} has precisely these properties. Namely, if we start with our self-correcting sequence
\begin{equation}
    S=\left[Z'\,X'^d\right]^{d-1}Z'\,\left[X'\,Z'^d\right]^{d-1}X'\,,
\end{equation}
and substitute in $X'=X$, then by just removing the copies of $X^d=I$, this simplifies to $\left[Z'^dX\right]^d$, which is precisely the sequence \eqref{eq:zcuring} which cures $Z'$ with an exact $X$. Similarly, the sequence $S$ cures $X'$ symmetrically --- when taking $Z'$ to be the exact $Z$, the copies of $Z^d$ resolve to the identity and one obtains a copy of equation \eqref{eq:xcuring}. This completes the proof that our sequence \eqref{eq:selfcorrectingexample} is self-correcting.\\

Essentially what is happening in our self-correcting sequence is that the curing sequences for $X'$ and $Z'$ are intertwined with one another. In this parent sequence, the last copy of $\left[ZX'^d\right]^{d-1}$ in the $X'$-curing sequence (and the first copy of $\left[XZ'^d\right]^{d}$ in the $Z'$-curing sequence) are expanded and padded with resolutions of the identity $X^d=Z^d=I$ --- which to first order ensures the sequence cures both types of errors simultaneously. 
This type of ``padding construction'' can be made to work based on any group with a $d$-dimensional irrep where the (projective) group decomposes into a product of orthogonal cyclic groups --- the Pauli group happens to be a minimal example of this.
We conjecture such self-correcting sequences can be made in other group structures, and we leave this to future work.
Even when using only using approximate generators $X'$ and $Z'$, self-correcting sequences are far from unique --- in fact our formal proof uses a slightly different version (see lemma \ref{lemma:XZd}) --- so we believe many other constructions exist, as long as they are built from elements which can generate the full group.\\

\vspace{-1em}
\subsection{Applications and discussion}

An inverse-free Solovay-Kitaev algorithm is of immediate interest in compiling unitaries from unconventional or noisy architectures or gate sets.
It also ensures that algorithmic runtimes are gate set-independent. It may also have nice geometric consequences --- for example our proof implies one can efficiently compile arbitrary rotations of a sphere using positive powers of two irrational rotations by non-collinear axes, which to the best of our knowledge was previously an open question. \\

We also note our result may have several applications in quantum complexity. 
For example, our result implies the true gate set independence of the definitions of exponentially precise complexity classes such as $\textsf{PostBQP}$ and $\textsf{QMA}_{\textsf{exp}}$.
These complexity classes often play an intermediate role in arguments in quantum complexity theory.
For example, as noted in \cite{bouland2017trading,bouland2016complexity,bouland2018complexity}, this gate set-independence might be useful in the classification of the computational power of non-universal quantum gate sets, since in these proofs one often makes use of certain ``postselection gadgets'' which are not intrinsically inverse closed. 
Furthermore, removing inverses from the gate set while maintaining expressivity might help lower the dimension in constructions used to study the complexity of translationally invariant local Hamiltonians \cite{bausch2017complexity1, bausch2017complexity2}.\\

Many open questions remain. First, we believe the {\em self-correcting} group sequences are a valuable object in their own right, which could be of interest to other areas of quantum information science. While we provide explicit constructions for the generalized Pauli groups, we conjecture that such sequences can be constructed with any approximations to a set of generators of a unitary irrep.\\

Second, there is the question if one can improve the resulting exponent in the poly-logarithmic asymptotic of the algorithm.
The standard, inverse-closed, Solovay-Kitaev algorithm has an exponent of $\approx 3.97$ \cite{dawson2005}; our inverse-free construction results in an exponent of $\approx 8.62$ for the qubit case\footnote{We note that there is a more sophisticated construction, relying on building nested group commutators, which we believe can reduce the exponent further. In particular the nested group commutator approach of \cite{kitaev2002classical, nielsen2010quantum} reduces the exponent from $\approx 3.97$ down to $2 + \nu$ (for any $\nu > 0$) in the inverse-closed case, and we conjecture this can be combined with our approach to slightly improve the exponent from roughly $\log_{3/2}(8d^2+1)$ to $\log_2 {8d^2}$, which we leave to future work.}, and an overall $O(\log d)$ scaling of the exponent with dimension. This logarithmic dependence of the exponent with dimension is intrinsic to any construction using group irreps, because the size of the group must be at least as large as the dimension of the matrix space in which it is irreducibly represented. New ideas are required to design algorithms which reduce the exponent further. Further motivation comes from the non-constructive proofs which show a realizable constant exponent of no more than three \cite{oszmaniec2020epsilon}, regardless of dimension --- while fundamental arguments place its lower bound at one \cite{nielsen2010quantum}. Algorithms which achieve this lower bound have been designed only for specific gate sets of practical interest (for example, \cite{ ross2014optimal, forest2015exact, bocharov2015efficient,kliuchnikov2015practical,parzanchevski2018super}). Reducing the exponent (and the hidden constants) in the case of arbitrary universal gate sets is crucial if gate set-agnostic constructions, such as the one proposed in our work, are to become practically useful. \\

Finally, it is a natural question to ask if our algorithm can be generalized to other Lie groups beyond $\text{SU}(d)$ or $\text{SL}(d,\mathbb{C})$ --- such as the more general class of connected Lie groups with perfect algebras, for which an inverse-closed Solovay-Kitaev algorithm was obtained in \cite{kuperberg2009hard}.

\subsection{Outline of paper}

In the rest of this paper we provide a formal proof of our inverse-free S-K theorem.
We first review the prior proofs of the S-K theorem in Section \ref{sec:review}, first covering the standard S-K algorithm \cite{kitaev1997quantum,dawson2005} in Section \ref{subsec:reviewinverses} and then the version with an irrep of \cite{sardharwalla2016universal,bouland2017trading} in Section \ref{subsec:reviewirrep}.
We then prove the $d=2$ case of our theorem in Section \ref{sec:qubit} followed by the case for general $\text{SU}(d)$ in Section \ref{sec:SU(d)}. Finally, in Section \ref{sec:SL(d,C)} we show how our techniques naturally extend to offer an inverse-free Solovay-Kitaev algorithm in $\text{SL}(d, \mathbb C)$.\\

\section{Background}
\label{sec:review}
\subsection{Notation and conventions}

In our analyses, we are interested in keeping track of terms to linear order in the error $\epsilon$. To this purpose, we employ the usual big-oh notation: we say that a term is $O(\epsilon^n)$ if there exists some $\epsilon_* > 0$ and some $C>0$ such that, for all $\epsilon \in [0,\, \epsilon_*)$, we have that the term in question is of magnitude less than $C\, \epsilon^n$.\\

To quantify the distance between operators we will use the operator norm as a distance function:
\begin{equation}
    D(U, V)=\norm{U-V}\equiv \sup_{\ip{\psi}{\psi}=1}\norm{(U-V)\ket{\psi}}_2\,.
\end{equation}
This norm is invariant under the applications of unitaries, namely $\norm{A}=\norm{AU}=\norm{UA}$ for any unitary $U$. In particular, this implies that we can rearrange the distance between two unitaries as a distance to the identity:
\begin{equation}
    \norm{U - V} = \norm{UV^\dagger - I}\,,
\end{equation}
a fact frequently used in developing the Solovay-Kitaev updates.
\\\\
Let us introduce some clarifying notation around the types of unitaries which appear in the algorithms in this work:
\begin{itemize}
    \item Normal typeface (e.g. $U$) is reserved for {\em abstract} representations of unitaries, while bold-faced notation (e.g. $\boldsymbol{U_n}$) represents a unitary implemented as a gate sequence. Such sequences can be composed in the normal sense of matrix multiplication: if $\boldsymbol{V_n}$ and $\boldsymbol{W_n}$ are two gate sequences, then their composition is $\boldsymbol{V_n\,W_n}$. If the gate set is inverse-closed, then taking the inverse $\boldsymbol{V_n^\dagger}$ of a sequence $\boldsymbol{V_n}$ is also allowed. We will use this notation when the context requires the distinction between a unitary as a classical representation, and a unitary as an implementable gate sequence. The goal of a Solovay-Kitaev-type algorithm is to go from an abstract representation $U$ of a unitary to an implementable sequence of gates $\boldsymbol{U_n}$ which approximates $U$ to within a certain error $\epsilon_n$. However, we can assume we can always access the classical representation $U_n$ of a gate sequence $\boldsymbol{U_n}$ at zero cost.
    
    \item We will use hats over unitaries to signify the precision of an approximation. A single hat means an approximation within an error of order $O(\epsilon)$; for example, $\overline{V^\dagger}$ would be an $\epsilon$-approximation to $V^\dagger$, or in other words, $\norm{V^\dagger - \overline{V^\dagger}}=O(\epsilon)$. Similarly, a double hat means an approximation to quadratic precision: $\norm{V^\dagger - \doverline{V^\dagger}}=O(\epsilon^2)$.
\end{itemize}

\subsection{Review of the Solovay-Kitaev algorithm with inverses \cite{kitaev1997quantum,dawson2005,nielsen2010quantum}}
\label{subsec:reviewinverses}

Here, we briefly review the standard Solovay-Kitaev algorithm, which assumes an inverse-closed gate set, based on the pedagogical review of Dawson and Nielsen \cite{dawson2005}. In a nutshell, the algorithm works recursively: at the $n$th step, it constructs a more accurate approximation of a target unitary $U$ based on an available $\epsilon_{n-1}$-approximation $\boldsymbol{U_{n-1}}$. The key step is to examine the difference unitary $UU_{n-1}^\dagger$, which is $\epsilon_{n-1}$ away from the identity, and to take advantage of the Lie algebra in the neighborhood of the identity to build an approximation of this difference to a stronger degree than $O(\epsilon_{n-1})$. This is realized by explicitly solving the following classical linear algebra problem:\\
\begin{tcolorbox}
\begin{lemma}[Approximate balanced group commutator \cite{dawson2005}]\label{lemma:ApproxGC}
    Given a matrix $\Delta \in \text{SU}(d)$ such that $\norm{\Delta - I}\leq \epsilon$, we can find unitaries $V$ and $W$ in $\text{SU}(d)$ such that their group commutator, defined as $VWV^\dagger W^\dagger$, approximates $\Delta$ to $\epsilon^{3/2}$ accuracy:
    \begin{equation}
        \norm{VWV^\dagger W^\dagger - \Delta} = O(\epsilon^{3/2})\,,
    \end{equation}
    and moreover $V$ and $W$ are both within a distance of order $\epsilon^{1/2}$ to the identity:
    \begin{equation}
        \norm{V - I} = O(\epsilon^{1/2})\,,\quad \norm{W - I}=O(\epsilon^{1/2})\,.
    \end{equation}
\end{lemma}
\end{tcolorbox}
\begin{proof}
We provide here a sketch of the argument based on \cite{dawson2005}. Any unitary in $\text{SU}(d)$ can be represented as $V=e^{iA}$ for some traceless, Hermitian $A$. Therefore, we can replace $V$, $W$, and $\Delta$ with $e^{iA}$, $e^{iB}$, and $e^{iH}$ respectively. Moreover, since we know $\Delta$, we can also find $H$ such that $\norm{H}=O(\epsilon)$. Performing the BCH expansion of the group commutator, we crucially have that the quadratic term in $A$ and $B$ cancels out and we are left with the next-to-leading contribution at third degree in $A$ and $B$:
    \begin{equation}
        VWV^\dagger W^\dagger = e^{iA}e^{iB}e^{-iA}e^{-iB} = I + [A,B] + O\left(\{A,B\}^3\right)\,.
    \end{equation}
Setting this equal to $\Delta=e^{iH}=I+iH+O(\epsilon^2)$ means that an equivalent problem is finding traceless, Hermitian matrices $A$ and $B$ of asymptotically equal norm $\norm{A}, \norm{B} = O(\norm{H}^{1/2})$ such that:
    \begin{equation}
        \label{eq:commutatorcondition}
        [A,B] = iH\,.
    \end{equation}
In \cite{dawson2005}, a simple solution to this classical linear algebra problem is proposed: first, move to the basis which is the Fourier-conjugate of the basis in which $H$ is diagonal. The traceless-ness of $H$ means that, in this basis, all diagonal elements of $H$ are zero. Take $B$ to be the diagonal matrix $B=\text{diag}\left(-\frac{d-1}{2}, -\frac{d-1}{2} + 1, \dots, \frac{d-1}{2}\right)$. Then our condition \eqref{eq:commutatorcondition} imposes that the elements of $A$ satisfy:
    \begin{equation}
        A_{jk} = \left\{\begin{array}{rr}
             \frac{iH_{jk}}{k-j}&\text{for }j\neq k  \\
             0 & \text{for }k=j
        \end{array}\right.\,.
    \end{equation}
    Therefore, $A$ and $B$ are both Hermitian and satisfy the desired condition \eqref{eq:commutatorcondition}. However, they are not yet balanced since we have that $\norm{A}\leq \sqrt{d\norm{H}}$ while $\norm{B}=\frac{d-1}{2}$. This is easily fixed by rescaling $A$ and $B$ appropriately to balance their norms. More sophisticated solutions to this problem are available which remove the $d$-dependence in the prefactors down to a constant \cite{kitaev2002classical}.
\end{proof}

In the Solovay-Kitaev routine at the $n$th step, we apply lemma \ref{lemma:ApproxGC} to construct a classical group commutator which approximates the difference unitary $UU_{n-1}^\dagger$ to a precision of order $O(\epsilon_{n-1}^{3/2})$. To this polynomially more precise approximation to $UU_{n-1}^\dagger$, we can simply append $\boldsymbol{U_{n-1}}$ to obtain a more precise approximation to $U$ itself, as desired. However, there is one more step: the construction in lemma \ref{lemma:ApproxGC} returns classical matrices $V$ and $W$, and we must translate them into gate sequences. The answer comes from the algorithm's recursive structure: just like we obtained a $\epsilon_{n-1}$-approximation to $U$ by calling the $(n-1)$th step of Solovay-Kitaev, we can do the same to the classical matrices $V$ and $W$. This returns the gate sequences $\boldsymbol{V_{n-1}}$ and $\boldsymbol{W_{n-1}}$ which are within $\epsilon_{n-1}$ of $V$ and $W$. At first, this approximation error seems larger than the $O(\epsilon_{n-1}^{3/2})$ precision offered by the group commutator construction. However, the ability to construct inverses exactly comes to the rescue in exactly canceling this approximation error at linear order. This effect is shown in the following lemma:\\
\begin{tcolorbox}
\begin{lemma}[Error cancelation in the approximate group commutator --- Lemma 1 in \cite{dawson2005}]\label{lemma:errorcancelGC}
    Suppose $\norm{V - V_{n-1}} \leq \epsilon_{n-1}$ and $\norm{W-W_{n-1}}\leq \epsilon_{n-1}$. Also, assume that $\norm{V-I}=O(\epsilon_{n-1}^{1/2})$ and $\norm{W-I}=O(\epsilon_{n-1}^{1/2})$. Then the group commutator in $V$, $W$ can be approximated by the group commutator in $V_{n-1}$, $W_{n-1}$ to a precision of order $O(\epsilon_{n-1}^{3/2})$:
    \begin{equation}
        \norm{VWV^\dagger W^\dagger - V_{n-1}W_{n-1}V^\dagger_{n-1}W^\dagger_{n-1}} = O(\epsilon_{n-1}^{3/2})\,.
    \end{equation}
\end{lemma}
\end{tcolorbox}
\begin{proof}
    The key idea is that exact inverses make terms linear in the error cancel out. Let us use the notation:
    \begin{equation}
        V_{n-1} = V + \Delta V\,,\quad W_{n-1} = W + \Delta W\,.
    \end{equation}
    Then unitarity means:
    \begin{align}
        V_{n-1}V_{n-1}^\dagger &= I \\
        (V+\Delta V)(V^\dagger + \Delta V^\dagger) &= I \\
        I + \Delta V V^\dagger +  V \Delta V^\dagger + \Delta V \Delta V^\dagger &= I\,,
    \end{align}
    in other words, the linear term in the error is in fact quadratic:
    \begin{equation}
        \label{eq:VDV}
        \Delta V V^\dagger + V \Delta V^\dagger  = - \Delta V \Delta V^\dagger = O(\epsilon_{n-1}^2)\,.
    \end{equation}
    Similarly for $W$:
    \begin{equation}
        \label{eq:WDW}
        \Delta W W^\dagger + W \Delta W^\dagger  = O(\epsilon_{n-1}^2)\,.
    \end{equation}
    With this in mind, let us expand the difference between the two group commutators and use the triangle inequality:
    \begin{align}
        \norm{VWV^\dagger W^\dagger - V_{n-1}W_{n-1}V^\dagger_{n-1}W^\dagger_{n-1}} &\leq \norm{\Delta V W V^\dagger W^\dagger + V \Delta W V^\dagger W^\dagger + VW\Delta V^\dagger W^\dagger + VWV^\dagger \Delta W^\dagger} + O(\epsilon_{n-1}^2) \\
        &\leq \norm{\Delta V W V^\dagger + V W \Delta V^\dagger} + \norm{\Delta W V^\dagger W^\dagger + WV^\dagger \Delta W^\dagger} + O(\epsilon_{n-1}^2)\,.
    \end{align}
    Each of these leading terms, which appear at first glance to be of order $\epsilon_{n-1}$, is in fact of order $\epsilon_{n-1}^{3/2}$. To see this, take the first term and expand $W$ around $I$, and again apply the triangle inequality:
    \begin{align}
        \norm{\Delta V W V^\dagger + V W \Delta V^\dagger} \leq \norm{\Delta V V^\dagger + V \Delta V^\dagger} + \norm{\Delta V(W - I)V^\dagger} + \norm{V(W- I)\Delta V^\dagger}\,.
    \end{align}
    Because of the inverse condition derived above in \eqref{eq:VDV}, the first term is quadratic in the error. Recall that, by assumption, $W-I=O(\epsilon_{n-1}^{1/2})$, which makes the next two terms of order $O(\norm{\Delta V}\norm{W-I})=O(\epsilon_{n-1}^{3/2})$ --- these are now the leading-order term in our difference. The same argument applies to the mirror term involving the inverse of $W$. Putting everything together, we get the stated asymptotic.
    \end{proof}

This shows that, when using the group commutator structure, the ability to construct precise inverses is crucial. When the gate set is inverse-closed, the exact inverses of any gate sequences can be natively produced as the reversed sequence of single-gate inverses. Putting everything together, we build the approximation of $U$ at the $n$th level as:
\begin{equation}
    \label{eq:UnSKdefinition}
    \boldsymbol{U_n} \equiv \boldsymbol{V_{n-1}}\boldsymbol{W_{n-1}}\boldsymbol{V_{n-1}^\dagger}\boldsymbol{W_{n-1}^\dagger}\boldsymbol{U_{n-1}}\,,
\end{equation}
which approximates $U$ at order $O(\epsilon_{n-1}^{3/2})$. We can see this by applying the triangle inequality and using the results of the above lemmas \ref{lemma:ApproxGC} and \ref{lemma:errorcancelGC}:
\begin{align}
    \label{eq:finalapproximationSK}
    \norm{U_n - U} &= \norm{V_{n-1}W_{n-1}V_{n-1}^\dagger W_{n-1}^\dagger U_{n-1} - U} \\
    &\leq \norm{{V_{n-1}}{W_{n-1}}{V_{n-1}^\dagger}{W_{n-1}^\dagger}U_{n-1} - VWV^\dagger W^\dagger U_{n-1}} + \norm{VWV^\dagger W^\dagger U_{n-1} - U} \\
    &=\norm{{V_{n-1}}{W_{n-1}}{V_{n-1}^\dagger}{W_{n-1}^\dagger}- VWV^\dagger W^\dagger} + \norm{VWV^\dagger W^\dagger - UU_{n-1}^\dagger}\\
    & \leq O(\epsilon_{n-1}^{3/2}) + O(\epsilon_{n-1}^{3/2}) = O(\epsilon_{n-1}^{3/2})\,.
\end{align}

Let us reassemble all the components above into a practical Solovay-Kitaev algorithm:
\\\\
\begin{algorithm}[H]
\setstretch{1.2}
  \caption{Standard Solovay-Kitaev with inverses \cite{dawson2005}}
  \label{alg:sk}
  \SetKwInput{kwInit}{Require}
  \kwInit{Gate set $\mathcal G$ containing its own exact inverses.}
  \Function{SK($U$, $n$)}{
    \tcp{Start the recursion at a basic approximation to $U$ with error $\epsilon_0$:}
    \If{$n=0$}{
        \Return{\FuncCall{$\epsilon_0$-net}{$U$}}\;
  }
    \Else{
    \tcp{Approximate $U$ to error $\epsilon_{n-1}$ by recursively calling the lower level of S-K:}
    $\boldsymbol{U_{n-1}} \assign \FuncCall{SK}{$U$, $n-1$}$\;
    \tcp{Solve the classical linear algebra problem according to the recipe in lemma \ref{lemma:ApproxGC}:}
    $V, W \assign \FuncCall{ApproximateGroupCommutator}{$U\,U_{n-1}^\dagger$}$\;
    \tcp{Approximate $V$, $W$ recursively within an error of $\epsilon_{n-1}$:}
    $\boldsymbol{V_{n-1}} \assign \FuncCall{SK}{$V$, $n-1$}$, \hspace{5pt}
    $\boldsymbol{W_{n-1}} \assign \FuncCall{SK}{$W$, $n-1$}$\;
    \tcp{Assemble the $n$th approximation to $U$ according to \eqref{eq:UnSKdefinition}:}
    \Return{$\boldsymbol{U_n}\equiv \boldsymbol{V_{n-1}}\boldsymbol{W_{n-1}}\boldsymbol{V_{n-1}^\dagger} \boldsymbol{W_{n-1}^\dagger} \boldsymbol{U_{n-1}}$}
    }
}
\end{algorithm}

\vspace{10pt}To analyze the asymptotics of this algorithm, note that the length of the sequence for $\boldsymbol{U_n}$ is five times the length of the sequences returned by the $(n-1)$th level of the algorithm, while the error decreases from $\epsilon_{n-1}$ to $O(\epsilon_{n-1}^{3/2})$. This defines the recursion in length and error:
\begin{align}
    \epsilon_n &\leq C\,\epsilon_{n-1}^{3/2}\label{eq:scalinginepsilon}\\
    \ell_n &= 5\,\ell_{n-1}\,,
\end{align}
for some constant $C>0$. Solving for the asymptotic of this recursion gives the well-known poly-logarithmic complexity of Solovay-Kitaev \cite{dawson2005}:
\begin{equation}
    \ell = O\left(\log^\gamma \epsilon^{-1}\right)\,,\quad \gamma=\frac{\log 5}{\log \frac32} \approx 3.97\,.
\end{equation}
An important aspect of S-K is the actual computational runtime of assembling the approximating sequence. This scales more slowly than the length of the sequence due to the redundancy in composing the sequence $\boldsymbol{U_n}$: the computation time is only spent on building the gate sequences $\boldsymbol{U_{n-1}}$, $\boldsymbol{V_{n-1}}$, and $\boldsymbol{W_{n-1}}$, after which the sequence \eqref{eq:UnSKdefinition} which makes up $\boldsymbol{U_n}$ can be assembled at no extra cost. We can then write a recursion for the runtime $t_n$ as $t_n = 3t_{n-1} + t_\text{GC}$, since the time is made up of three calls to the S-K subroutine, plus the time to solve the linear algebra problem of the approximate group commutator (lemma \ref{lemma:ApproxGC}), assumed to be a constant $t_\text{GC}$. Therefore, the runtime of S-K will scale as $t = O(\log^{2.7}\epsilon^{-1})$ where the exponent is $\log 3 / \log\frac{3}{2} \approx 2.7$.\\

One important final comment: in algorithm \ref{alg:sk}, as well as in all the other algorithms in this work, we assume that we can start the Solovay-Kitaev recusion at level $n=0$ by guaranteeing an $\epsilon_0$-approximation to our target matrix $U$. The role of an S-K-type algorithm is to take sequences which can approximate any unitary in $\text{SU}(d)$ to an appropriate precision and use them to construct exponentially more precise sequences. Here, ``appropriate'' refers to the upper bound on how large $\epsilon_0$ is allowed to be; based on \eqref{eq:scalinginepsilon}, we have that $\epsilon_0\leq1/C^2$ for the recursion to work. In general, this means a requirement that $\epsilon_0 \leq 1/\text{poly}(d)$ where $d$ is the dimension of the Hilbert space. This is where the universality of the gate sequence factors in: we assume that we can build enough sequences of an appropriate length such that they densely cover $\text{SU}(d)$ with an $\epsilon_0$-net, and then query this $\epsilon_0$-net as an oracle to initialize our algorithms.

\subsection{Review of Solovay-Kitaev with an irrep \cite{sardharwalla2016universal,bouland2017trading}}
\label{subsec:reviewirrep}
In this section, we outline the advances made by \cite{sardharwalla2016universal, bouland2017trading} in reformulating the Solovay-Kitaev algorithm to accommodate a different structure in the gate set. Instead of assuming that the gate set contains the inverse of every element within, this section assumes that the only requirement on the gate set is that it contains an exact irreducible representation of any finite group. In other words, the assumption is that there exists an irreducible representation (irrep) $R:G \to \text{SU}(d)$ such that $\{R(g)\,:\,g\in G\}\subset \mathcal G$, where $\mathcal G$ is our universal gate set which densely generates $\text{SU}(d)$. The idea of \cite{sardharwalla2016universal, bouland2017trading} is to use the irrep to explicitly construct a way to produce operator inverses --- which we call {\em inverse factories} --- which will be correct to quadratic order in $\epsilon$. This is possible by applying a group twirl (lemma \ref{lemma:twirl} below) to project operators onto the identity. The quadratic precision of the resulting inverse factories (lemma \ref{lemma:inversefactoryirrep} below) is enough to use them instead of the usual exact inverses in the standard Solovay-Kitaev algorithm \ref{alg:sk}, and still maintain the S-K poly-logarithmic asymptotics. The constructions in this section are summarized in the form of algorithm \ref{alg:skirrep}.\\

In the functioning of the standard Solovay-Kitaev algorithm \ref{alg:sk}, we took advantage of being able to build exact inverses in the case of an inverse-closed gate set. Suppose, however, that we could only build approximate inverses for the operators $V_{n-1}$ and $W_{n-1}$ at every step. How good would such an approximate inverse factory need to be? The answer is: at least as good as the $O(\epsilon_{n-1}^{3/2})$ precision resulting from approximating the group commutator. We can state this explicitly as the following lemma:\\
\begin{tcolorbox}
\begin{lemma}[Approximating the group commutator]\label{lemma:ApproxGCapproxInverses}
    For two unitaries $V$ and $W$, assume we have two $\epsilon_n$-approximations $V_n$ and $W_n$, meaning that $\norm{V - V_n}=O(\epsilon_n)$ and $\norm{W - W_n}=O(\epsilon_n)$. Additionally assume that $V$ and $W$ are close to the identity to order $O(\epsilon_n^{1/2})$, namely $\norm{V-I}=O(\epsilon_n^{1/2})$ and $\norm{W-I}=O(\epsilon_n^{1/2})$. Finally, assume we have precise approximations to the inverses of $V_n$ and $W_n$ such that $\norm{\doverline{V_n^\dagger} - V_n^\dagger}=O(\epsilon_n^{3/2})$ and $\norm{\doverline{W_n^\dagger} - W_n^\dagger}=O(\epsilon_n^{3/2})$. Then the group commutator of $V$ and $W$ can still be approximated to order $O(\epsilon_n^{3/2})$:
    \begin{equation}
        \norm{VWV^\dagger W^\dagger - V_nW_n\doverline{V_n^\dagger}\,\doverline{W_n^\dagger}} = O(\epsilon_n^{3/2})\,.
    \end{equation}
\end{lemma}
\end{tcolorbox}
\begin{proof}
    This is a simple extension of lemma \ref{lemma:errorcancelGC}. Expanding $\doverline{V_n^\dagger}$ around $V_n^\dagger$ (and similarly for $\doverline{W_n^\dagger}$) and applying the triangle inequality, we obtain:
    \begin{align*}
        \norm{VWV^\dagger W^\dagger - V_n W_n \doverline{V_n^\dagger}\, \doverline{W_n^\dagger}} \leq &  \norm{VWV^\dagger W^\dagger - V_n W_n V_n^\dagger W_n^\dagger} + \\
        & +\norm{\doverline{V_n^\dagger} - V_n^\dagger} + \norm{\doverline{W_n^\dagger} - W_n^\dagger} + \norm{\doverline{V_n^\dagger} - V_n^\dagger}\norm{\doverline{W_n^\dagger} - W_n^\dagger}
    \end{align*}
    Since by assumption the inverses are precise to order $O(\epsilon_n^{3/2})$, the terms after the first come in precisely this order, or higher in the case of the final cross-term. Finally, lemma \ref{lemma:errorcancelGC} tells us that the first term is of order $O(\epsilon_n^{3/2})$. Therefore, the error in the inverses is at the same order and the group commutator is overall approximated to order $O(\epsilon_n^{3/2})$ just like in the standard Solovay-Kitaev.
\end{proof}

As it turns out, we will not saturate this $O(\epsilon^{3/2})$ upper-bound on the necessary precision in constructing inverses. The existing methods for producing inverses using exact group irreps, which we discuss in this section, as well as our construction for the inverse-free case, both produce inverses to an even tighter precision, of order $O(\epsilon^2)$. It is an open question whether closing this gap can lead to a lower exponent in the algorithm's asymptotics, and we leave this to future work.\\

Next, we review the contributions of \cite{sardharwalla2016universal, bouland2017trading} in showing how to use an exact group irrep in the gate set in order to construct an {\em inverse factory} for arbitrary gate sequences in $\text{SU}(d)$. We start with a basic fact about how twirling over an irrep acts as a projection onto the identity:\\

\begin{tcolorbox}
\begin{lemma}[Twirling over irreps projects onto the identity \cite{bouland2017trading}]\label{lemma:twirl}
    Let $R:G \to \text{SU}(d)$ be an irreducible unitary representation of a finite group $G$.
    Then twirling an operator $A$ over the elements of the irrep projects the operator $A$ onto the identity:
    \begin{equation}
        \sum_{g\in G} R(g)\,A\,R(g)^\dagger = \frac{\abs{G}}{d}\tr(A)\,I\,,
    \end{equation}
    where $\abs{G}$ is the number of elements in the group $G$.
\end{lemma}
\end{tcolorbox}
\begin{proof}
    The matrix element of an irrep satisfy the Schur orthogonality property \cite{serre1977linear}, which is mathematically the same as the property of being a unitary one-design:
    \begin{equation}
        \sum_{g\in G} \mel{i}{R(g)}{j}\mel{k}{R(g)^\dagger}{l} = \frac{\abs{G}}{d}\delta_{i,l}\delta_{j,k}\,.
    \end{equation}
    Applying this fact to the $i,j$ matrix element of the twirl gives us:
    \begin{align}
        \sum_{g\in G} \mel{i}{R(g)\,A\,R(g)^\dagger}{j} &= \sum_{g\in G}\sum_{k,l=0}^{d-1}\mel{i}{R(g)}{k}\mel{k}{A}{l}\mel{l}{R(g)^\dagger}{j} \\
        &= \sum_{k,l=0}^{d-1}\mel{k}{A}{l}\sum_{g\in G} \mel{i}{R(g)}{k}\mel{l}{R(g)^\dagger}{j} \\
        &= \sum_{k,l=0}^{d-1}\mel{k}{A}{l}\,\frac{\abs{G}}{d}\,\delta_{i,j}\delta_{k,l} \\
        &=\frac{\abs{G}}{d}\sum_{k=0}^{d-1}\mel{k}{A}{k}\,\delta_{i,j} \\
        &= \frac{\abs{G}}{d}\tr(A)\,\mel{i}{I}{j}\,.
    \end{align}
\end{proof}

Now let us use this insight to show how an exact irrep in $\text{SU}(d)$ can be used to focus operators onto the identity to quadratic precision. The following lemma improves on the constant obtained in a similar result in Appendix A of \cite{bouland2017trading}:\\
\begin{tcolorbox}
\begin{lemma}[Focusing on the identity with an irrep \cite{bouland2017trading}]\label{lemma:onedesignfocusing}
    Let $R:G\to \text{SU}(d)$ be an irrep of a finite group $G$, and define the (multiplicative) twirl of an operator over the elements of $R(G)$ as:
    \begin{equation}
        f_R(A) \equiv R(g_1)\,A\,R(g_1)^\dagger\,R(g_2)\,A\,R(g_2)^\dagger\dots R(g_\abs{G})\,A\,R(g_\abs{G})^\dagger = \prod_{g \in G}R(g)\,A\,R(g)^\dagger\,.
    \end{equation}
    Let $I' \in \text{SL}(d, \mathbb C)$ be an $\epsilon$-approximation to the identity, namely $\norm{I'-I}\leq \epsilon$. Then twirling $I'$ over the irrep $R$ approximates the identity to quadratic order:
    \begin{equation}
        f_{R}(I') = I + O(\epsilon^2)\,.
    \end{equation}
\end{lemma}
\end{tcolorbox}
\begin{proof}
    Denote $I' = I + \Delta$, where by assumption $\Delta$ is of order $O(\epsilon)$. Let us first bring in the well-known expansion of the determinant around the identity, which to linear order is:
    \begin{equation}
        \det(I') = \det(I + \Delta) = 1 + \tr(\Delta) + O(\epsilon^2)\,.
    \end{equation}
    Combining this with the fact that $\det I' = 1$ (because $I'\in \text{SL}(d, \mathbb C)$), we must have that $\tr(\Delta)=O(\epsilon^2)$.
    \\\\
    This fact is essential in showing that, when expanding the twirl, the linear term in $\Delta$ is actually quadratic in $\epsilon$:
    \begin{align}
      f_\mathcal{R}(I') &= \prod_{g \in G} R(g)\,(I + \Delta)\,R(g)^\dagger \\
      &= I + \sum_{g \in G} R(g)\,\Delta\,R(g)^\dagger + O(\epsilon^2) \\
      &= I + \frac{\abs{G}}{d}\tr(\Delta)\,I + O(\epsilon^2) \\
      &= I + O(\epsilon^2),
    \end{align}
    which proves the claim by applying the calculation in lemma \ref{lemma:twirl} on the first-order term in the expansion. Note that we do not need the operator $I'$ to necessarily be unitary. We have only used the facts that $I'$ is $\epsilon$-close to $I$, and that its determinant is one, so we can more generally take $I'\in\text{SL}(d,\mathbb C)$.
\end{proof}

Finally, we can use this fact about the twirl construction to build a quadratically-precise {\em inverse factory} with our exact irrep:\\
\begin{tcolorbox}
    \begin{lemma}[Inverse factory with an irrep \cite{bouland2017trading}]\label{lemma:inversefactoryirrep}
        Let $R:G \to \text{SU}(d)$ be an irrep of a finite group $G$. Let $V$ be an operator in $\text{SU}(d)$, and assume we have access to an approximate inverse of $V$ in the form of $\overline{V^\dagger} \in \text{SU}(d)$ such that $\norm{V^\dagger - \overline{V^\dagger}}\leq \epsilon$. Then the following construction provides an inverse to $V$ which is precise to order $O(\epsilon^2)$:
        \begin{equation}
            \label{eq:inversefactoryirrep}
            \doverline{V^\dagger} \equiv \left[\prod_{g\in G\setminus \{\text{id}\}}R(g) \overline{V^\dagger}V R(g)^\dagger\right] \overline{V^\dagger}\,,
        \end{equation}
        meaning that $\norm{V\doverline{V^\dagger} - I}=O(\epsilon^2)$.
    \end{lemma}
\end{tcolorbox}
\begin{proof}
    This is an immediate consequence of the group twirl result in lemma \ref{lemma:onedesignfocusing}. Notice that the group twirl there works for any ordering of the group elements. In particular, it works for an ordering of the group elements in which the identity group element $\text{id}$ is last in the sequence. Applying lemma \ref{lemma:onedesignfocusing} with such an order to twirl the operator $\overline{V^\dagger}V$, which by assumption is $\epsilon$-close to the identity and in $\text{SU}(d)$, gives us:
    \begin{equation}
        \left[\prod_{g\in G \setminus \{\text{id}\}} R(g)\overline{V^\dagger}VR(g)^\dagger\right]\overline{V^\dagger}V = I + O(\epsilon^2)\,.
    \end{equation}
    Therefore, everything in this sequence excluding the final instance of $V$ serves as an inverse of $V$ to order $O(\epsilon^2)$, as needed.
\end{proof}

Equipped with this inverse factory, it is now possible to adjust the standard Solovay-Kitaev algorithm \ref{alg:sk} for the case in which the gate set contains an exact irrep, but no other additional structure. We do this by using the inverse factory in \eqref{eq:inversefactoryirrep} to replace the exact inverses in the group commutator. By lemma \ref{lemma:ApproxGCapproxInverses}, this is more than enough to maintain the group commutator asymptotics, since the error in the inverse factories is $O(\epsilon_{n-1}^2)$, which is even smaller than the $O(\epsilon_{n-1}^{3/2})$ requirement of lemma \ref{lemma:ApproxGCapproxInverses}.\\

Algorithm \ref{alg:skirrep} assembles all these components explicitly. Note that the length of the sequence resulting from the $n$th step of the algorithm now scales as:
\begin{equation}
    \ell_{n} = (4\abs{G}+1)\ell_{n-1} + 2(\abs{G}-1)\,,
\end{equation}
since the sequence includes $4\abs{G}+1$ instances of sequences obtained from the $(n-1)$th level of S-K, plus $2\abs{G}-2$ raw elements of the gate set in the form of the irrep operators. The scaling in the error $\epsilon$ is unchanged from \eqref{eq:scalinginepsilon}, hence the sequence length of Solovay-Kitaev with an irrep asymptotes as $\ell=O\left(\log^{\log(4\abs{G}+1)/\log(3/2)}\epsilon^{-1}\right)$. Recall that the size of an irrep is at least the dimension of the matrix space, i.e. $\abs{G}\geq d^2$ --- therefore the exponent of the poly-logarithmic scaling will itself scale as $O(\log d)$.
\begin{algorithm}
\setstretch{1.2}
  \caption{Solovay-Kitaev with an exact irrep \cite{bouland2017trading}}
  \label{alg:skirrep}
  \SetKwInput{kwInit}{Require}
  \kwInit{Universal gate set $\mathcal G$ containing an exact irrep $\boldsymbol{R}:G \to \text{SU}(d)$.}
  \Function{SK($U$, $n$)}{
    \tcp{Start the recursion at a basic approximation to $U$ with error $\epsilon_0$:}
    \If{$n=0$}{
        \Return{\FuncCall{$\epsilon_0$-net}{$U$}}\;
  }
    \Else{
    
    \tcp{Approximate $U$ to error $\epsilon_{n-1}$ by recursively calling the lower level of S-K:}
    $\boldsymbol{U_{n-1}} \assign \FuncCall{SK}{$U$, $n-1$}$\;
    \tcp{Solve the group commutator problem according to the recipe in lemma \ref{lemma:ApproxGC}:}
    $V, W \assign \FuncCall{ApproximateGroupCommutator}{$U\,U_{n-1}^\dagger$}$\;
    \tcp{Obtain sequences for $V$, $W$ recursively within an error of $\epsilon_{n-1}$:}
    $\boldsymbol{V_{n-1}} \assign \FuncCall{SK}{$V$, $n-1$}$, \hspace{5pt}
    $\boldsymbol{W_{n-1}} \assign \FuncCall{SK}{$W$, $n-1$}$\;
    \tcp{Obtain sequences for the inverses of $\boldsymbol{V_{n-1}}$, $\boldsymbol{W_{n-1}}$ recursively within an error of $\epsilon_{n-1}$:}
    $\boldsymbol{\overline{V_{n-1}^\dagger}} \assign \FuncCall{SK}{$V_{n-1}^\dagger$, $n-1$}$, \hspace{5pt}
    $\boldsymbol{\overline{W_{n-1}^\dagger}} \assign \FuncCall{SK}{$W_{n-1}^\dagger$, $n-1$}$\;
    \tcp{Construct quadratically precise sequences for the inverses to $\boldsymbol{V_{n-1}}$ and $\boldsymbol{W_{n-1}}$ using \eqref{eq:inversefactoryirrep}:}
    $\doverline{\boldsymbol{V_{n-1}^\dagger}} \assign \left[\prod_{g\in G,\,g\neq\text{id}}\boldsymbol{R}(g)\boldsymbol{\overline{V_{n-1}^\dagger}}\boldsymbol{V_{n-1}}\boldsymbol{R}(g)^\dagger\right] \boldsymbol{\overline{V_{n-1}^\dagger}}$ \;
    
    $\doverline{\boldsymbol{W_{n-1}^\dagger}} \assign \left[\prod_{g\in G,\,g\neq\text{id}}\boldsymbol{R}(g)\boldsymbol{\overline{W_{n-1}^\dagger}}\boldsymbol{W_{n-1}}\boldsymbol{R}(g)^\dagger\right] \boldsymbol{\overline{W_{n-1}^\dagger}}$ \;
    
    \tcp{Return the $O(\epsilon_{n-1}^{3/2})$-precise sequence:}
    
    \Return{$\boldsymbol{U_n}\equiv \boldsymbol{V_{n-1}}\boldsymbol{W_{n-1}}\doverline{\boldsymbol{V_{n-1}^\dagger}}\, \doverline{\boldsymbol{W_{n-1}^\dagger}} \boldsymbol{U_{n-1}}$}\;
    }
}
\end{algorithm}

\section{Inverse-Free Solovay-Kitaev on a Qubit}\label{sec:qubit}

In this section we develop an inverse-free Solovay-Kitaev algorithm, with no assumption on the structure of the universal gate set $\mathcal G$, for the two-dimensional case of unitaries acting on a qubit. This is a significant illustrative example which, despite its simplicity, contains all the key conceptual steps found in the generic $\text{SU}(d)$ case, while still being intuitively accessible. We start by describing the algebra of approximate Pauli operators. Due to the Lie algebra of the special unitary group, an approximate Pauli unitary will differ from the true Pauli by an error which is quadratically suppressed in the direction of the true Pauli unitary itself. This opens the possibility of combining multiple approximate Paulis in sequences which, due to the projective structure of the group, will suppress errors in multiple directions. In fact, we find that it is possible to build {\em self-correcting sequences} which suppress errors in all directions at linear order. In lemma \ref{lemma:selfinverse}, we construct such sequences which quadratically approximate the identity with only $\epsilon$-precise Paulis. Finally, we show how it is possible to turn such self-correcting sequences into generic inverse factories, which take an $\epsilon$-approximate inverse to arbitrary unitaries, and return an improved $\epsilon^2$-approximation to the inverse. This quadratic precision in building inverses is fit for use in a Solovay-Kitaev routine instead of the exact inverses in the standard S-K algorithm \ref{alg:sk} --- leading to the inverse-free qubit Solovay-Kitaev algorithm \ref{alg:ifskqubit}.

\subsection{The algebra of approximate Pauli operators}\label{sec:preliminaries}

Let the exact Pauli operators be defined in the standard way as:
\begin{equation}
    X\equiv \begin{pmatrix} 0 & 1 \\ 1 & 0\end{pmatrix}\,,\quad Y\equiv \begin{pmatrix} 0 & -i \\ i & 0\end{pmatrix}\,,\quad Z\equiv \begin{pmatrix} 1 & 0 \\ 0 & -1\end{pmatrix}\,.
\end{equation}
Note that the Paulis $X, Y, Z$, together with the identity $I$, form a basis for all $2\times 2$ matrices over the complex numbers. We start by using this basis to study the structure of errors in the neighborhood of the Pauli operators.\\

\begin{tcolorbox}
\begin{lemma}[Suppression of error directions]\label{lemma:suppresion} If $X'$ is an $\epsilon$-approximation to $X$ such that $\norm{X' - X}\leq \epsilon$, then the error $X'-X$ is suppressed in the direction of $X$. Specifically, there are complex scalars $x_0$, $x_2$, $x_3$ of order $O(\epsilon)$ such that:
\begin{equation}
    \label{eq:suppression}
    X' = X + x_0 I + x_2 Y + x_3 Z + O(\epsilon^2)\,.
\end{equation}
\end{lemma}
\end{tcolorbox}
\begin{proof}
    Since $X'$ and $X$ are both $2\times 2$ special unitaries, then the unitary connecting them is guaranteed to be special (i.e. of determinant one), and therefore we can write:
    \begin{equation}
        X' = X\, e^{iaX + ibY + icZ}\,,
    \end{equation}
    for some real $a,b,c$ which must be of order $O(\epsilon)$. Note that the lack of a component along $I$ in the exponent is a direct result of the traceless Lie algebra of $\text{SU}(2)$. Expanding the exponential, we have:
    \begin{equation}
        X' = X \left(I + iaX + ibY + icZ + O(\epsilon^2)\right) = X + iaI - bZ + cY + O(\epsilon^2)\,,
    \end{equation}
    which is precisely the form in \eqref{eq:suppression}.
\end{proof}

Clearly, similar results apply to $\epsilon$-approximations to each of the Paulis so we can generally write:
\begin{align}
    X' &= X + x_0 I + x_2 Y + x_3 Z + O(\epsilon^2) \label{eq:Xapprox}\\
    Y' &= Y + y_0 I + y_1 X + y_3 Z + O(\epsilon^2) \label{eq:Yapprox}\\
    Z' &= Z + z_0 I + z_1 X + z_2 Y + O(\epsilon^2) \label{eq:Zapprox}\\
    I' &= I + j_1 X + j_2 Y + j_3 Z + O(\epsilon^2) \label{eq:Iapprox}\,,
\end{align}
where $x_k, y_k, z_k, j_k$ are of order $O(\epsilon)$.\\

Next, we demonstrate the ability to build {\em self-correcting sequences} made of $\epsilon$-approximate Paulis, which surprisingly self-focus onto the identity to order $O(\epsilon^2)$:\\

\begin{tcolorbox}
\begin{lemma}[Self-correcting sequences from approximate Paulis]\label{lemma:selfinverse}
If $\norm{X' - X}\leq \epsilon$ and $\norm{Y' - Y}\leq \epsilon$, then the construction:
\begin{equation}
    J_2(X', Y') \equiv X' Y' X' Y'^2 X' Y' X'
\end{equation}
is an $\epsilon^2$-approximation to the identity, i.e. $\norm{J_2(X', Y') - I}=O(\epsilon^2)$.
\end{lemma}
\end{tcolorbox}
\begin{proof}
    We apply the result of lemma \ref{lemma:suppresion} and carry out the calculation. In particular, using \eqref{eq:Xapprox} and \eqref{eq:Yapprox}:
    \begin{align*}
        X'Y' & = iZ + (x_2 + y_1)I + (ix_3 + y_0)X + (x_0 -iy_3)Y + O(\epsilon^2)\\
        Y'X' & = -iZ + (x_2 + y_1)I + (-ix_3 + y_0)X + (x_0 + iy_3)Y + O(\epsilon^2)\,.
    \end{align*}
    When squaring an approximate Pauli, only the leading errors aligned with $I$ and the nearest Pauli survive to leading order. In the case of the above, we are close to $\pm iZ$ but have no leading $Z$ error component so when squaring we are left with errors only in the $Z$ direction at order $\epsilon$:
    \begin{align*}
        X'Y'X'Y' &= -I + 2i(x_2 + y_1)Z + O(\epsilon^2) \\
        Y'X'Y'X' &= -I - 2i(x_2 + y_1)Z + O(\epsilon^2)\,.
    \end{align*}
    We now have two mirror sequences which, to leading order, are of the form $-I \pm \epsilon \Delta + O(\epsilon^2)$, so when multiplied the errors cancel to leading order, hence:
    \begin{equation}
        J_2(X', Y') = X'Y'X'Y'\times Y'X'Y'X' = I + O(\epsilon^2)\,,
    \end{equation}
    as desired.
\end{proof}

Note that in we can peel off an $X'$ from one of the two ends of the $J_2$ sequence in lemma \ref{lemma:selfinverse} to obtain a second-order approximation to $X'^\dagger$, such as:
\begin{equation}
    \doverline{X'^\dagger}\equiv X'Y'X'Y'^2X'Y'=X'^\dagger + O(\epsilon^2)\,.
\end{equation}
Note however that $\norm{\doverline{X'^\dagger}-X}$ is still of order $O(\epsilon)$. By an argument based on the cyclic permutation symmetry of Pauli matrices, we can find analogous sequences involving the other Pauli matrices.
    
\subsection{Inverse factories and the qubit algorithm}

Next, we show how assembling approximate Paulis into self-correcting sequences is enough to construct a second-order `inverse factory' for arbitrary $2\times 2$ unitaries:\\

\begin{tcolorbox}
    \begin{lemma}[Quadratic inverse factory in $\text{SU}(2)$]\label{lemma:inversefactoryqubit} Assume we have access to Pauli approximations $X'$, $Y'$, $Z'$  such that $\norm{X' - X}\leq \epsilon$, $\norm{Y'-Y}\leq \epsilon$, and $\norm{Z' - Z}\leq \epsilon$. Assume that, given a unitary $V$, we have an  $\epsilon$-approximate inverse $\overline{V^\dagger}$ such that $\norm{\overline{V^\dagger} - V^\dagger}\leq \epsilon$. Then the construction:
\begin{equation}
    \doverline{V^\dagger} \equiv X' \left(\overline{V^\dagger}\,V\right) Y' X' \left(\overline{V^\dagger}\,V\right) Y'^2 X' \left(\overline{V^\dagger}\,V\right) Y' X' \,\overline{V^\dagger} 
\end{equation}
is an $\epsilon^2$-approximation to $V^\dagger$, i.e. $\norm{\doverline{V^\dagger}V - I} = O(\epsilon^2)$.
\end{lemma}
\end{tcolorbox}
\begin{proof}
    The proof is extremely quick in light of lemma \ref{lemma:selfinverse}, which showed that the $J_2(X', Y')$ sequence approximates the identity to quadratic precision in $\epsilon$, as long as $X'$ and $Y'$ are $O(\epsilon)$-approximations to the true Pauli $X$ and $Y$ operators.\\
    
    Note that since $\overline{V^\dagger}$ is an $\epsilon$-approximate inverse to $V$, then the product $\overline{V^\dagger}V$ is an $\epsilon$-approximation to the identity. We can combine this with an $\epsilon$-approximation to a Pauli $X$ operator to obtain another $\epsilon$-approximate $X$ operator, namely:
    \begin{equation}
        X'\,\overline{V^\dagger}\,V = (X + O(\epsilon))(I + O(\epsilon)) = X + O(\epsilon)\,.
    \end{equation}
    Therefore, $X'\,\overline{V^\dagger}\,V$ is an $\epsilon$-approximation to $X$ and can be used instead of the original $X'$ in the $J_2$ construction of lemma \ref{lemma:selfinverse}. Specifically, this means that the following sequence is self-correcting onto the identity:
    \begin{equation}
        \label{eq:selfcorrectingXYVVqubit}
        J_2(X'\,\overline{V^\dagger}\,V,\, Y') = X' \left(\overline{V^\dagger}\,V\right) Y' X' \left(\overline{V^\dagger}\,V\right) Y'^2 X' \left(\overline{V^\dagger}\,V\right) Y' X' \left(\overline{V^\dagger}\,V\right) = I + O(\epsilon^2)\,.
    \end{equation}
    Denoting everything in this sequence except the last instance of $V$ by $\doverline{V^\dagger}$ means that $\doverline{V^\dagger}V=I + O(\epsilon^2)$, and therefore $\doverline{V^\dagger}$ is an $\epsilon^2$-approximation to the inverse of $V$.
\end{proof}

It is worth pausing to understand how the construction in lemma \ref{lemma:inversefactoryqubit} is in fact performing precisely the type of group twirls used in \cite{sardharwalla2016universal, bouland2017trading} which we discussed in Section \ref{subsec:reviewirrep}. Denote $I'\equiv \overline{V^\dagger}V \equiv I + \Delta I$ as the unitary which is $\epsilon$-close to the identity. Similarly, let $X' = X + \Delta X$ and $Y' = Y + \Delta Y$, with the understanding that all errors $\Delta I$, $\Delta X$, and $\Delta Y$ are of order $O(\epsilon)$. When expanding the self-correcting construction in \eqref{eq:selfcorrectingXYVVqubit}, we get:
\begin{align}
    J_2(X'I', Y') &= X' I' Y' X' I' Y'^2 X' I' Y' X' I' \\
    &= I + \sum_{P \in \{I, X, Y, Z\}} P\left(\Delta I + \Delta X\,X + \Delta Y\, Y\right)\,P + O(\epsilon^2)\\
    &= I + O(\epsilon^2)\,.
\end{align}
Here, the term linear in the errors $\Delta I$, $\Delta X$, $\Delta Y$ is quadratically suppressed because it is precisely the type of group twirl (in this case, over the Pauli group) which matches the conditions of lemmas \ref{lemma:twirl} and \ref{lemma:onedesignfocusing}. This works because $\Delta I$, $\Delta X\,X$, and $\Delta Y\,Y$ are matrices of size $O(\epsilon)$ which are traceless to order $O(\epsilon^2)$ --- a consequence of lemma \ref{lemma:suppresion} and its corollaries \eqref{eq:Xapprox}, \eqref{eq:Yapprox}, and \eqref{eq:Iapprox}.\\

To sum up, lemma \ref{lemma:inversefactoryqubit} provides an explicit inverse factory for any operator $V \in \text{SU}(d)$. Most importantly, it only requires components which are precise up to order $O(\epsilon)$ (the approximate Paulis and the approximate inverse to $V$), and it returns an inverse sequence which is precise to order $O(\epsilon^2)$. This is exactly the tool we need to substitute instead of the exact inverses in Solovay-Kitaev, since the $O(\epsilon)$-precise components can be obtained from the lower level recursion of S-K, while the quadratic precision in inverse-building is enough to maintain the standard S-K asymptotics according to lemma \ref{lemma:ApproxGCapproxInverses}. Therefore, we are now ready to formulate the inverse-free Solovay-Kitaev algorithm for a qubit in the form of algorithm \ref{alg:ifskqubit}.

\begin{algorithm}
 \setstretch{1.2}
  \caption{Inverse-free Solovay-Kitaev on a qubit}
    \label{alg:ifskqubit}
  \SetKwInput{kwInit}{Require}
  \kwInit{Any universal gate set $\mathcal G \subset \text{SU}(2)$.}
  \Function{IFSK($U$, $n$)}{
    \tcp{Start the recursion by a basic $\epsilon_0$-approximation to $U$ based on universality of $\mathcal G$:}
    \If{$n=0$}{
        \Return{\FuncCall{$\epsilon_0$-net}{$U$}}\;
  }
    \Else{
        \tcp{Obtain $\epsilon_{n-1}$-approximation sequence to $U$ recursively:}
        $\boldsymbol{U_{n-1}} \assign \FuncCall{IFSK}{$U$, $n-1$}$\;
        \tcp{Solve the group commutator problem based on lemma \ref{lemma:ApproxGC}:}
        $V, W \assign \FuncCall{ApproximateGroupCommutator}{$U\,U_{n-1}^\dagger$}$\;
        \tcp{Obtain the sequences which $\epsilon_{n-1}$-approximate $V$, $W$, and the inverses, recursively:}
        $\boldsymbol{V_{n-1}} \assign \FuncCall{IFSK}{$V$, $n-1$}$, \hspace{5pt}
        $\boldsymbol{W_{n-1}} \assign \FuncCall{IFSK}{$W$, $n-1$}$\;
        $\boldsymbol{\overline{V_{n-1}^\dagger}} \assign \FuncCall{IFSK}{$V_{n-1}^\dagger$, $n-1$}$, \hspace{5pt}
        $\boldsymbol{\overline{W_{n-1}^\dagger}} \assign \FuncCall{IFSK}{$W_{n-1}^\dagger$, $n-1$}$\;
        \tcp{Obtain sequences which $\epsilon_{n-1}$-approximate the Paulis (this step can be precomputed and stored):}
        $\boldsymbol{X_{n-1}} \assign \FuncCall{IFSK}{$X$, $n-1$}$, \hspace{5pt}
        $\boldsymbol{Y_{n-1}} \assign \FuncCall{IFSK}{$Y$, $n-1$}$\;
        \tcp{Use the inverse factory in lemma \ref{lemma:inversefactoryqubit} to the obtain $\epsilon_{n-1}^2$-precise inverse sequences for $\boldsymbol{V_{n-1}}$ and $\boldsymbol{W_{n-1}}$:}
        $\boldsymbol{\doverline{V_{n-1}^\dagger}} \assign \boldsymbol{X_{n-1} \overline{V_{n-1}^\dagger} V_{n-1} Y_{n-1} X_{n-1} \overline{V_{n-1}^\dagger} V_{n-1} Y_{n-1}^2 X_{n-1} \overline{V_{n-1}^\dagger} V_{n-1} Y_{n-1} X_{n-1} \overline{V_{n-1}^\dagger}}$\;
        $\boldsymbol{\doverline{W_{n-1}^\dagger}} \assign \boldsymbol{X_{n-1} \overline{W_{n-1}^\dagger} W_{n-1} Y_{n-1} X_{n-1} \overline{W_{n-1}^\dagger} W_{n-1} Y_{n-1}^2 X_{n-1} \overline{W_{n-1}^\dagger} W_{n-1} Y_{n-1} X_{n-1} \overline{W_{n-1}^\dagger}}$\;
        \tcp{Finally, return the sequence which approximates $U$ to order $O(\epsilon_{n-1}^{3/2})$:}
        \Return{$\boldsymbol{U_{n}} \equiv \boldsymbol{V_{n-1}} \boldsymbol{W_{n-1}}\boldsymbol{\doverline{V_{n-1}^\dagger}}\,\boldsymbol{\doverline{W_{n-1}^\dagger}}\,\boldsymbol{U_{n-1}}$}
    }
}
\end{algorithm}

\vspace{10pt}\begin{tcolorbox}
\begin{theorem}[Complexity of IFSK on a qubit]\label{thrm:ifskqubit}
    Algorithm \ref{alg:ifskqubit} solves the approximation problem to qubit gate $U$ within an error of $\epsilon$ with gate sequences of length $O\left(\log^\gamma \epsilon^{-1}\right)$ where the exponent is $\gamma = \frac{\log 33}{\log 3/2}\approx 8.62$. 
\end{theorem}
\end{tcolorbox}
\begin{proof}
   To obtain the asymptotic complexity, note that if the gate length to obtain an $\epsilon_{n-1}$-approximation to any unitary is $\ell_{n-1}$, then the prescriptions for the quadratically-precise inverses $\boldsymbol{\doverline{V_{n-1}^\dagger}}$ and $\boldsymbol{\doverline{W_{n-1}^\dagger}}$ have length $15\ell_{n-1}$ each. Putting everything together, we get that the resulting sequence for $\boldsymbol{U_n}$ has overall length:
    \begin{equation}
        \ell_{n} = 33 \ell_{n-1}\,.
    \end{equation}
    Combining this with the typical S-K scaling of the error \eqref{eq:scalinginepsilon} results in the stated asymptotic. Note however, that the computational runtime of determining the sequence is significantly shorter. Assuming we precompute and store the approximate Paulis and their self-corrected inverses, then each inverse-free S-K step calls the lower level recursively precisely 5 times: to obtain $\epsilon_{n-1}$-sequences for each of $U, V, W, V_{n-1}^\dagger,$ and $W_{n-1}^\dagger$. This results in an algorithm runtime of $t=O\left(\log^{3.97}\epsilon^{-1}\right)$ with an exponent of $\frac{\log 5}{\log \frac{3}{2}}\approx 3.97$.
\end{proof}

\section{Generalizing to $\text{SU}(d)$}\label{sec:SU(d)}
In this section we generalize the results of Section \ref{sec:qubit} to the special unitary group in an arbitrary dimension $d\geq 2$. For this purpose, we promote the qubit Pauli operators to the generalized Pauli group (sometimes called the Weyl group). The steps towards the inverse-free S-K algorithm follow those of the qubit case exactly, starting with finding the necessary self-correcting sequences in $\text{SU}(d)$ and using them to construct a precise inverse factory for arbitrary unitaries. Just like in the qubit case, this ability is then used to produce the inverses required by the Solovay-Kitaev routine. This workflow is implemented in algorithm \ref{alg:usk}, the inverse-free Solovay-Kitaev on $\text{SU}(d)$.

\subsection{Definitions and preliminaries}

In $d$ dimensions, we can use a standard generalization of the Pauli unitaries, in which $X$ is promoted to a {\em shift operator} and $Z$ becomes the {\em clock operator}:
\begin{align}
    \omega &\equiv e^{\frac{2\pi i}{d}} \\
    Z &\equiv \sum_{j=0}^{d-1} \omega^j \dyad{j}{j} \\
    X &\equiv \sum_{j=0}^{d-1} \dyad{j+1\Mod d}{j}
\end{align}
Let us furthermore, when convenient, use the notation:
\begin{equation}
    \sigma(n,m) \equiv X^nZ^m\,.
\end{equation}
With these generators, the resulting group has the following properties:
\begin{itemize}
    \item $X^d=Z^d=I$. Therefore, we will always understand the powers of $X$ and $Z$ modulo $d$, for example $\sigma(-k,0)=\sigma(d-k,0)$.
    \item All group elements are (up to a global phase) of the form $\sigma(n,m)=X^n Z^m$ with $n,\,m\in \{0,1,\dots,d-1\}$.
    \item The commutation rule is $ZX=\omega XZ$.
    \item The element inverse is $\left(X^nZ^m\right)^\dagger = \omega^{nm}X^{d-n}Z^{d-m}$.
    \item The set $\{X^nZ^m\}_{n,m=0}^{d-1}$ is a projective irrep of the generalized Pauli group. According to lemma \ref{lemma:onedesignfocusing}, an exact implementation of this set of operators would be enough for an inverse factory in $\text{SU}(d)$. 
\end{itemize}

Next, we show how to generalize lemma \ref{lemma:suppresion} to the $d$-dimensional case: namely, that the error of an approximate Pauli is quadratically suppressed in the direction of the exact Pauli itself. This extension occurs quite naturally because, just like in the $d=2$ case, the $d^2$ generalized Pauli operators of the form $X^nZ^m$ also form a basis for all $d\times d$ matrices. Moreover, all of the generalized Paulis except the identity are traceless --- which makes this basis a natural fit for the traceless algebra of $\text{SU}(d)$. We are therefore led to the following result:\\

\begin{tcolorbox}
\begin{lemma}\label{lemma:suppressionSU(d)}
    Let $\sigma'(n,m)$ be an $\epsilon$-approximation to the generalized Pauli operator $\sigma(n,m)=X^nZ^m$, such that $\norm{\sigma'(n,m) - X^nZ^m}\leq \epsilon$. Then, to leading order in $\epsilon$, the error is orthogonal to $X^nZ^m$, namely:
    \begin{equation}
        \sigma'(n,m) = X^nZ^m + \sum_{(a,b)\neq(0,0)}\theta_{a,b}X^{n+a}Z^{m+b} + O(\epsilon^2)\,,
    \end{equation}
    where $\theta_{a,b}$ are complex numbers of size $O(\epsilon)$. Here, the notation means that the sum is over all pairs of indices except the $(0,0)$ one: $(a,b) \in \{0,1,\dots,d-1\}^2\setminus\{(0,0)\}$.
\end{lemma}
\end{tcolorbox}
\begin{proof}
Since all the $\sigma(n,m)$ form a set of $d^2$ linearly independent matrices, they can serve as a basis for $d\times d$ matrices. Therefore, we can once again take advantage of the traceless algebra of $\text{SU}(d)$ to express an $\epsilon$-approximation to $\sigma(n,m)$ as:
\begin{align}
    \sigma'(n,m) &= \sigma(n,m)e^{\sum_{(a, b) \neq (0, 0)}\theta_{a,b}\sigma(a, b)} \\
    &= \sigma(n, m)\left(I + \sum_{(a,b)\neq (0, 0)}\theta_{a, b}\sigma(a, b) + O(\epsilon^2)\right) \\
    &= \sigma(n, m) + \sum_{(a, b) \neq (0, 0)}\omega^{ma}\theta_{a, b}\sigma(n + a, m + b) + O(\epsilon^2) \\
    &= \sigma(n, m) + \sum_{(a, b) \neq (n, m)} \omega^{m(a-n)}\theta_{a-n, b-m}\sigma(a, b) + O(\epsilon^2)\,.\label{eq:sudsuppression}
\end{align}
Here, all $\theta_{a,b}$ are of order $O(\epsilon)$ --- this shows how to extend lemma \ref{lemma:suppresion} to dimension $d$ in that the error along the nearest $\sigma(n,m)$ is quadratically suppressed. Note that unitarity imposes further constraints on $\theta_{a,b}$ (i.e. that the exponent is anti-Hermitian) but we will never make use of this condition.
\end{proof}

\subsection{Self-correcting sequences in $\text{SU}(d)$}

In this section we show how to generalize the self-correcting qubit construction in lemma \ref{lemma:selfinverse} to an arbitrary dimension $d$. While we choose to study a certain instance of self-correcting sequence of the shortest length of $2d^2$, the choice is far from unique. Our formal proof of the self-correcting property relies on explicit calculation; for a more intuitive, informal discussion of the mechanism behind self-correcting sequences, see the explanation in Section \ref{subsec:selfcorrectdiscussion}.\\

\begin{tcolorbox}
\begin{lemma}[Self-corrected sequences of generalized $X$ and $Z$]\label{lemma:XZd}
    For dimension $d\geq 2$, let the generalized Pauli group be $\sigma(n,m)\equiv X^n Z^m$, where $X$ and $Z$ are the shift and clock operator in $\text{SU}(d)$ respectively. Define the function:
    \begin{equation}
        J_d(A, B)\equiv \left[A^d B\right]^{d-1} A \left[B^d A\right]^{d-1} B\,.
    \end{equation}
    If we have access to approximations $X'$ and $Z'$ to $X$ and $Z$ such that $\norm{X' - X}\leq\epsilon$ and $\norm{Z' - Z}\leq\epsilon$, then the following sequences self-focus on the identity:
    \begin{align}
        J_d(X', Z') &= I + O(\epsilon^2) \label{eq:XZconstruction}\\
        J_d(Z', X') &= I + O(\epsilon^2)\,.\label{eq:ZXconstruction}
    \end{align}
\end{lemma}
\end{tcolorbox}
\begin{proof}
    The proof follows from explicit calculation. Here we outline some key steps in showing that the terms linear in $\epsilon$ cancel out in the sequence $J_d(Z',X')$.
    \\\\
    Using the notation $\sigma(n,m)=X^nZ^m$ and the commutation rule $ZX=\omega XZ$, we find the multiplication rule: $\sigma(n_1,m_1)\sigma(n_2,m_2)=\omega^{n_2m_1}\sigma(n_1+n_2, m_1+m_2)$. We start from the finding \eqref{eq:sudsuppression} that the error in $\sigma(n,m)$ is suppressed in the direction of $\sigma(n,m)$ to explicitly expand:
    \begin{align}
        Z' &= \sigma'(0,1) = \sigma(0,1) + \sum_{(a,b)\neq(0,0)} x_{a,b}\sigma(a,b+1) + O(\epsilon^2) \\
        X' &= \sigma'(1,0) = \sigma(1,0) + \sum_{(a,b)\neq(0,0)} z_{a,b}\sigma(a+1,b) + O(\epsilon^2)\,,
    \end{align}
    where $x_{a,b}$ and $z_{a,b}$ are complex numbers of order $O(\epsilon)$ which describe the degrees of freedom in the errors. Our goal is to show that the terms linear in the $x_{a,b}$ and $z_{a,b}$ vanish in the proposed sequence, regardless of the specific values of these errors. We can explicitly compute:
    \begin{align}
        \sigma'(0,1)^d &= I + \sum_{(a,b)\neq(0,0)} x_{a,b}\sigma(a,b)\sum_{k=0}^{d-1}\omega^{ka} + O(\epsilon^2)=I + d\sum_{b=1}^{d-1}x_{0,b}\sigma(0,b) + O(\epsilon^2) \\
        \sigma'(1,0)^d &= I + \sum_{(a,b)\neq(0,0)}z_{a,b}\sigma(a,b)\sum_{k=0}^{d-1}\omega^{kb} +O(\epsilon^2) = I + d\sum_{a=1}^{d-1}z_{a,0}\sigma(a,0) + O(\epsilon^2)\,,
    \end{align}
    where the identity $\sum_{k=0}^{d-1}\omega^{kN}=d\,\delta_{0,\,N \Mod d}$ is used to show that $d-1$ error terms out of $d^2-1$ survive to leading order. At the next step we have:
    \begin{align}
        \left[\sigma'(0,1)^d\sigma'(1,0)\right]^{d-1} &= \sigma(-1,0) +\left( d\sum_{b=1}^{d-1}x_{0,b}\omega^{b}\sigma(-1,b) + \sum_{(a,b)\neq(0,0)}z_{a,b}\sigma(a-1, b)\right)\sum_{k=0}^{d-2}\omega^{kb}+O(\epsilon^2) \\
        \left[\sigma'(1,0)^d\sigma'(0,1)\right]^{d-1} &= \sigma(0,-1) + \left(
            d\sum_{a=1}^{d-1}z_{a,0}\sigma(a,-1) + \sum_{(a,b)\neq(0,0)}x_{a,b}\sigma(a,b-1)
        \right)\sum_{k=0}^{d-2}\omega^{ka}+O(\epsilon^2)\,.
    \end{align}
    Finally, when combining all the elements together, we can collect the terms linear in $\epsilon$ corresponding to the $x_{a,b}$ and, separately, $z_{a,b}$:
    \begin{align}
        J_d(\sigma'(0,1),\,\sigma'(1,0)) =& \left[\sigma'(0,1)^d\sigma'(1,0)\right]^{d-1}\sigma'(0,1)\left[\sigma'(1,0)^d\sigma'(0,1)\right]^{d-1}\sigma'(1,0) \\
        =& I + \sum_{(a,b)\neq (0,0)}\omega^b x_{a,b}\sigma(a,b)\left[ d\,\delta_{a,0}\omega^b\sum_{k=0}^{d-2}\omega^{bk} + \omega^a\sum_{k=0}^{d-2}\omega^{ka}+1\right] + \label{eq:xbrackets}\\
        &+ \sum_{(a,b)\neq(0,0)} z_{a,b}\sigma(a,b)\left[d\,\delta_{b,0}\omega^a\sum_{k=0}^{d-2}\omega^{ka} + \omega^b\sum_{k=0}^{d-2}\omega^{bk} + 1\right] + O(\epsilon^2)\,.\label{eq:zbrackets}
    \end{align}
    It is enough to show that the terms in the brackets vanish for any $(a,b)\in\{0,1,\dots,d-1\}^2\setminus \{(0,0)\}$. We can rearrange the standard root identity $\sum_{k=0}^{d-1}\omega^{kN}=d\,\delta_{0,\,N\Mod d}$ as:
    \begin{equation}
        \sum_{k=0}^{d-2}\omega^{ka} = (d-1)\delta_{a,0} - \omega^{-a}(1-\delta_{a,0})\quad\text{ for any }a\in\{0,1,\dots,d-1\}\,.
    \end{equation}
    Using this, we can write the bracket in \eqref{eq:xbrackets} which multiplies $x_{a,b}$ as:
    \begin{align}
        d\,\delta_{a,0}\omega^b\sum_{k=0}^{d-2}\omega^{bk} + \omega^a\sum_{k=0}^{d-2}\omega^{ka}+1 &= d\delta_{a,0}\delta_{b,0}\omega^b\left(d-1+\omega^{-b}\right) + \delta_{a,0}\omega^{a}\left(\omega^{-a}-1\right)=0\,,
    \end{align}
    since the first term is nonzero only if both $a$ and $b$ are simultaneously zero, which is an instance we do not sum over in \eqref{eq:xbrackets}, and the second term is zero both when $a=0$ (in which case $\omega^{-a}-1$ would be zero), and when $a\neq 0$ (in which case $\delta_{a,0}$ would be zero). Similarly, the term associated with $z_{a,b}$ vanishes by the same arguments. Therefore, regardless of the exact values of the errors $x_{a,b}$ and $z_{a,b}$, their contribution vanishes to linear order, and the sequence self-corrects onto the identity to quadratic order in $\epsilon$. Note that the $J_d$ sequence has length $2d^2$ in terms of $X'$ and $Z'$.
\end{proof}

\subsection{Inverse factories and an inverse-free Solovay-Kitaev algorithm}

Just like in the qubit case, the construction of self-correcting sequences of approximate Pauli generators is extremely useful in constructing precise inverses to any operator in $\text{SU}(d)$ for which we already have an approximate inverse. This is precisely the framework needed in a Solovay-Kitaev routine, where we only have access to $\epsilon$-approximate implementations of arbitrary unitaries from recursively calling the function --- and in which we need to engineer operations with these $\epsilon$-approximate elements which are accurate to a higher order in $\epsilon$.\\

\begin{tcolorbox}
        \begin{lemma}[Inverse factory in $\text{SU}(d)$]\label{lemma:inversesd}
        Assume we have $\epsilon$-approximations to the generalized Pauli generators $X$ and $Z$ in the form of $X'$ and $Z'$. Specifically, we have that $\norm{X' - X}\leq \epsilon$ and $\norm{Z' - Z}\leq \epsilon$. Furthermore, assume that for an operator $V$, we have an $\epsilon$-approximate inverse $\overline{V^\dagger}$, such that $\norm{\overline{V^\dagger} - V^\dagger} = \norm{\doverline{V^\dagger}V - I}\leq \epsilon$. Then we can obtain a quadratically precise inverse of $V$ in the form of the sequence:
        \begin{equation}
            \label{eq:inversefactory}
            \doverline{V^\dagger} \equiv \left[X'^d \left(Y'\overline{V^\dagger}V\right)\right]^{d-1}X'\left[\left(Y'\overline{V^\dagger}V\right)^d X'\right]^{d-1}Y'\overline{V^\dagger}\,.
        \end{equation}
        Specifically, $\norm{\doverline{V^\dagger}V - I} = \norm{\doverline{V^\dagger} - V^\dagger} = O(\epsilon^2)$. Moreover, the length of the inverse sequence is $4d^2-1$ in terms of the $X'$, $Z'$, $V$, and $\overline{V^\dagger}$ components.
    \end{lemma}
\end{tcolorbox}
\begin{proof}
    This simple proof relies on the same observation as the $d=2$ case laid out in lemma \ref{lemma:inversefactoryqubit}. Since $Y'$ is an $\epsilon$-approximation to $Y$, and since $\overline{V^\dagger}V$ is an $\epsilon$-approximation to the identity, then their product $Y'\overline{V^\dagger}V$ is an $\epsilon$-approximation to $Y$. This means that this product can be substituted instead of $Y'$ in the $J_d$ self-correcting sequences obtained in lemma \ref{lemma:XZd}. Specifically, since we showed that $J_d(X',\, Y')$ is a self-correcting sequence which approximates the identity quadratically, then so is $J_d(X', \,Y'\overline{V^\dagger}V)$:
    \begin{equation}
        J_d(X',\,Y'\overline{V^\dagger}V)=\left[X'^d \left(Y'\overline{V^\dagger}V\right)\right]^{d-1}X'\left[\left(Y'\overline{V^\dagger}V\right)^d X'\right]^{d-1}\left(Y'\overline{V^\dagger}V\right) = I + O(\epsilon^2)\,.
    \end{equation}
    Finally, this sequence has an instance of $V$ on the rightmost position, which is extremely convenient in building the inverse to $V$. Peeling off this $V$ from the side of the sequence gives us the $\epsilon^2$-approximate inverse  $\doverline{V^\dagger}$, since $\doverline{V^\dagger}V = I + O(\epsilon^2)$. Finally, the sequence for the inverse has a total length of $4d^2-1$ in terms of the $X'$, $Y'$, $V$, and $\overline{V^\dagger}$ building blocks.
\end{proof}

\begin{algorithm}
 \setstretch{1.2}
  \caption{Inverse-free Solovay-Kitaev in $\text{SU}(d)$.}
    \label{alg:usk}
  \SetKwInput{kwInit}{Require}
  \kwInit{Any universal gate set $\mathcal G \subset \text{SU}(d)$.}
  \Function{IFSK($U$, $n$)}{
    \tcp{Start the recursion at precision $\epsilon_0$ by using universality of $\mathcal G$:}
    \If{$n=0$}{
        \Return{\FuncCall{$\epsilon_0$-net}{$U$}}\;
  }
    \Else{
        \tcp{Obtain sequence which $\epsilon_{n-1}$-approximates $U$ recursively:}
        $\boldsymbol{U_{n-1}} \assign \FuncCall{IFSK}{$U$, $n-1$}$\;
        \tcp{Solve for the group commutator according to lemma \ref{lemma:ApproxGC}:}
        $V, W \assign \FuncCall{ApproximateGroupCommutator}{$U\,U_{n-1}^\dagger$}$\;
        \tcp{Obtain sequence which $\epsilon_{n-1}$-approximate $V$, $W$, and the corresponding inverses, recursively:}
        $\boldsymbol{V_{n-1}} \assign \FuncCall{IFSK}{$V$, $n-1$}$, \hspace{5pt}
        $\boldsymbol{W_{n-1}} \assign \FuncCall{IFSK}{$W$, $n-1$}$\;
        $\boldsymbol{\overline{V_{n-1}^\dagger}} \assign \FuncCall{IFSK}{$V_{n-1}^\dagger$, $n-1$}$, \hspace{5pt}
        $\boldsymbol{\overline{W_{n-1}^\dagger}} \assign \FuncCall{IFSK}{$W_{n-1}^\dagger$, $n-1$}$\;
        \tcp{Obtain sequences for the generalized Pauli group generators $X$ and $Z$ to precision $O(\epsilon_{n-1})$, recursively (this step can be precomputed and stored):}
        $\boldsymbol{X_{n-1}}\assign \FuncCall{IFSK}{$X$, $n-1$}$,\hspace{5pt} $\boldsymbol{Z_{n-1}}\assign \FuncCall{IFSK}{$Z$, $n-1$}$\;
        \tcp{Use the recipe in \eqref{eq:inversefactory} to build inverse sequences for $\boldsymbol{V_{n-1}}$ and $\boldsymbol{W_{n-1}}$, precise to order $O(\epsilon_{n-1}^2)$:}
        $\boldsymbol{\doverline{V_{n-1}^\dagger}} \assign \boldsymbol{\left[X_{n-1}^d \left(Y_{n-1}\overline{V_{n-1}^\dagger}V_{n-1}\right)\right]^{d-1}X_{n-1}\left[\left(Y_{n-1}\overline{V_{n-1}^\dagger}V_{n-1}\right)^d X_{n-1}\right]^{d-1}Y_{n-1}\overline{V_{n-1}^\dagger}}$\;
        $\boldsymbol{\doverline{W_{n-1}^\dagger}} \assign \boldsymbol{\left[X_{n-1}^d \left(Y_{n-1}\overline{W_{n-1}^\dagger}W_{n-1}\right)\right]^{d-1}X_{n-1}\left[\left(Y_{n-1}\overline{W_{n-1}^\dagger}W_{n-1}\right)^d X_{n-1}\right]^{d-1}Y_{n-1}\overline{W_{n-1}^\dagger}}$\;
        \Return{$\boldsymbol{U_n}\equiv \boldsymbol{V_{n-1}\,W_{n-1}\,\doverline{V_{n-1}^\dagger}\,\doverline{W_{n-1}^\dagger}\,U_{n-1}}$}
    }
}
\end{algorithm}

We are now in a position to formulate an inverse-free Solovay-Kitaev algorithm for arbitrary dimension in the form of algorithm \ref{alg:usk}. Note that the steps involving obtaining approximate group operators at every level $n$ only need to be performed once; we can store the result and use the same set of group operators over multiple uses of the algorithm for different target unitaries $U$.\\

\begin{tcolorbox}
\begin{theorem}[Complexity of inverse-free Solovay-Kitaev]\label{thm:usk}
    Algorithm \ref{alg:usk} constructs gate sequences which approximate arbitrary unitaries $U$ in $\text{SU}(d)$, of lengths which scale asymptotically as $\ell = O(\log^{\gamma_d} \epsilon^{-1})$. The exponent is $\gamma_d=\frac{\log(8d^2 + 1)}{\log (3/2)}=\Theta(\log d)$.
\end{theorem}
\end{tcolorbox}
\begin{proof}
    Lemma \ref{lemma:ApproxGCapproxInverses} again applies, showing that that being able to build quadratically-precise inverses for $\boldsymbol{V_{n-1}}$ and $\boldsymbol{W_{n-1}}$ guarantees that the final result will be of order $O(\epsilon_{n-1}^{3/2})$. In other words, we have that the sequence length and the error at every step $n$ of the algorithm follow the recursion:
    \begin{align}
        \epsilon_n &\leq C\,\epsilon_{n-1}^{3/2}\\
        \ell_n &= (8d^2 + 1)\ell_{n-1}\,.
    \end{align}
    Here, we have used that the two inverse factories each require $4d^2-1$ subcomponents of length $\ell_{n-1}$, as calculated in Lemma \ref{lemma:inversesd}. This recursion results in the stated asymptotic. Just like in the qubit case discussed in theorem \ref{thrm:ifskqubit}, the computational runtime of the algorithm scales more slowly than the length of the sequences. When precomputing and storing the approximate Pauli group sequences and their quadratic inverses, the runtime depends only on the five recursive calls to the S-K subroutines. Therefore, up to constant terms related to matrix operations, the runtime recursion is simply $t_n=5t_{n-1}$, which leads to an asymptotic for the runtime of $t=O\left(\log^{3.97}\epsilon^{-1}\right)$ since the exponent is $\frac{\log 5}{\log\frac{3}{2}}\approx 3.97$.
\end{proof}

\section{Inverse-free Solovay-Kitaev in $\text{SL}(d,\mathbb C)$}\label{sec:SL(d,C)}

Finally, in this section we describe how our techniques are enough to formulate an inverse-free Solovay-Kitaev algorithm for the special linear group over complex numbers $\text{SL}(d,\mathbb C)$. The key observation is that our inverse factory in lemma \ref{lemma:inversesd}, which produces operator inverses to precision $O(\epsilon^2)$, applies equally well to operators in $\text{SL}(d,\mathbb C)$, not just to the special use case of $\text{SU}(d)$ elaborated in Section \ref{sec:SU(d)}.\\

An inverse-closed Solovay-Kitaev algorithm for $\text{SL}(d, \mathbb C)$ was developed by Aharonov, Arad, Eban, and Landau in Appendix B of  \cite{aharonov2007polynomial}. A more general inverse-closed algorithm, which applies to any connected Lie group with a perfect Lie algebra (and therefore also to $\text{SU}(d,\mathbb C)$) was developed by Kuperberg in \cite{kuperberg2009hard}. Just like the standard $\text{SU}(d)$ S-K algorithm \ref{alg:sk}, both these extensions rely on using group commutators, which retain their useful approximation properties described in Section \ref{sec:review}. Similar to the standard Solovay-Kitaev, the inverses needed in the group commutators are assumed to be exact, and natively implemented by the inverse-closed gate set. To make use of group commutators in the inverse-free case, we again need an `inverse factory': a method to obtain $\epsilon^2$-precise inverses for operators in $\text{SL}(d, \mathbb C)$, starting from only an $\epsilon$-precise components. Similar to the $\text{SU}(d)$ case in Section \ref{sec:SU(d)}, such an inverse factory can be used in place of the exact inverses in the group commutators, and guarantee that the polylogarithmic asymptotic of the algorithm is maintained. \\

In fact, the inverse factory developed for $\text{SU}(d)$ in Section \ref{sec:SU(d)} can naturally be extended to the case of $\text{SL}(d, \mathbb C)$. This is due to the fact that the constructions used in building precise inverses only make use of the {\em special} property (i.e. that the determinant is one). First, the special property suppresses the error parallel to the generalized Pauli (lemma \ref{lemma:suppressionSU(d)}), which remains true even if the $\epsilon$-approximate Pauli is now in $\text{SL}(d, \mathbb C)$. In turn, this fact is enough to construct the self-correcting sequences in lemma \ref{lemma:XZd}. Furthermore, the action of a group twirl to produce an inverse (lemma \ref{lemma:onedesignfocusing}) was already formulated on $\text{SL}(d, \mathbb C)$. Therefore, our framework naturally extends to $\text{SL}(d, \mathbb C)$ to offer an inverse-free Solovay-Kitaev algorithm. More generally, following the example of \cite{kuperberg2009hard}, we expect our explicit construction to extend to any connected Lie group with a perfect algebra which contains an irreducible representation of $\text{SU}(d)$ as a subgroup.\\

\subsection*{Acknowledgments}

We thank Greg Kuperberg for helpful discussions. This material is based upon work supported in part by the U.S. DOE Office of Science under Award Number DE-SC0020266.

\end{document}